\documentclass[
  aps,
  prapplied,
  twocolumn,
  superscriptaddress,
  longbibliography
]{revtex4-2}

\usepackage[utf8]{inputenc}
\usepackage[T1]{fontenc}
\usepackage{amsmath, amssymb, amsthm}
\usepackage{graphicx}
\usepackage{xcolor}
\usepackage{float} 
\usepackage{comment}

\usepackage{braket}

\usepackage{tikz}
\usepackage{tikz-3dplot}
\usetikzlibrary{arrows.meta, positioning, shapes, decorations.pathreplacing}

\tikzset{
  dctrlslash/.style={
    draw,
    circle,
    minimum size=5pt,
    inner sep=0pt,
    line width=0.4pt,
    path picture={\draw[line width=0.8pt] (path picture bounding box.north west) -- (path picture bounding box.south east);}
  },
  dprimegate/.style={
    draw,
    circle,
    minimum size=18pt, 
    inner sep=0pt,
    line width=0.6pt,
    label=center:$\scriptstyle >D'$, 
  }
}

\usepackage{quantikz}

\usepackage{hyperref}

\newcommand{\K}{\mathrm{k}}
\newcommand{\dprime}{\gate[style={dprimegate}]{}}
\newcommand{\dctrlslash}{\gate[style={dctrlslash}]{}}

\makeatletter
\AtBeginDocument{%
  \def\bibsection{%
    \par
    \begingroup
      \baselineskip26\p@
      \bib@device{\columnwidth}{122.75\p@}%
    \endgroup
    \nobreak\@nobreaktrue
    \addvspace{19\p@}%
    \par
  }%
}
\makeatother
\begin{document}

\title{Block encoding the 3D heterogeneous Poisson equation with application to fracture flow}

\author{Austin Pechan}
\email{agp@berkeley.edu}
\affiliation{Theoretical Division, Los Alamos National Laboratory, Los Alamos, New Mexico 87545, USA}

\author{John Golden}
\email{golden@lanl.gov}
\affiliation{CCS-3, Los Alamos National Laboratory, Los Alamos, New Mexico 87545, USA}

\author{Daniel O'Malley}
\affiliation{EES-16, Los Alamos National Laboratory, Los Alamos, New Mexico 87545, USA}

\begin{abstract}
Quantum linear system (QLS) algorithms offer the potential to solve large-scale linear systems exponentially faster than classical methods. However, applying QLS algorithms to real-world problems remains challenging due to issues such as state preparation, data loading, and efficient information extraction. In this work, we study the feasibility of applying QLS algorithms to solve discretized three-dimensional heterogeneous Poisson equations, with specific examples relating to groundwater flow through geologic fracture networks. We explicitly construct a block encoding for the 3D heterogeneous Poisson matrix by leveraging the sparse local structure of the discretized operator. While classical solvers benefit from preconditioning, we show that block encoding the system matrix and preconditioner separately does not improve the effective condition number that dominates the QLS runtime. This differs from classical approaches where the preconditioner and the system matrix can often be implemented independently. Nevertheless, due to the structure of the problem in three dimensions, the quantum algorithm achieves a runtime of $O(N^{2/3} \ \text{polylog } N \cdot \log(1/\epsilon))$, outperforming the best classical methods (with runtimes of $O(N \log N \cdot \log(1/\epsilon))$) and offering exponential memory savings. These results highlight both the promise and limitations of QLS algorithms for practical scientific computing, and point to effective condition number reduction as a key barrier in achieving quantum advantages.

\end{abstract}
\maketitle
\section{Introduction}

The three-dimensional (3D) heterogeneous Poisson equation is a foundational model in science and engineering. This partial differential equation (PDE) describes steady-state diffusion processes in media with spatially varying properties. Applications range from modeling fluid dynamics in porous media and heat transfer in inhomogeneous materials to simulating electrical currents in biological tissues. While the underlying physics is often well-understood, solving these equations becomes computationally challenging when the medium's properties, such as permeability or conductivity, vary across many length scales.

For classical computers, the primary bottleneck in solving these problems is memory. Accurately capturing the system's behavior requires a numerical discretization fine enough to resolve the small-scale variations in material properties. For large domains, this leads to linear systems of equations that far exceed the memory capacity of classical supercomputers. Standard simplification techniques like coarse-graining or homogenization, which average over fine-scale details, often fail because they cannot capture critical long-range effects, such as percolation, that depend on the full, multiscale structure of the medium. Consequently, many important physical systems remain beyond the reach of full-resolution classical simulation.

Quantum linear system (QLS) algorithms offer a potential path forward. Recent algorithms have achieved a runtime complexity of $O(\kappa \log(1/\epsilon))$, which is information-theoretically optimal in the effective condition number $\kappa$ and error tolerance $\epsilon$~\cite{costa2021optimal, dalzell2024}. This scaling suggests an exponential advantage over classical methods for certain classes of problems. However, this theoretical promise is contingent on satisfying a strict set of requirements related to data loading, state preparation, and information extraction. This paper moves beyond theoretical potential to rigorously assess the circumstances under which QLS algorithms could deliver a practical quantum advantage, particularly in the case of the 3D heterogeneous Poisson equation. Our key contributions are:
\begin{itemize}
    \item An explicit description of a block encoding of the 3D heterogeneous Poisson matrix (discretization of $\nabla \cdot (\K(\mathbf{r}) \nabla h(\mathbf{r}))$).
    \item A proof that separately block encoding a matrix and its preconditioner can never improve the effective condition number of the system, and thus never improve the QLS runtime.
    \item A modest quantum speed-up of $O(N^{2/3} \text{ polylog } N \cdot \log(1/\epsilon))$, in comparison to classical algorithms that scale as $O(N \log N \cdot \log(1/\epsilon))$, for solving discretized 3D heterogeneous Poisson equations.
    \item An explicit example of our algorithm as applied to the problem of liquid flowing through geologic fracture networks, systematically addressing state preparation, data loading, condition number, and information extraction for this real-world problem.
\end{itemize}

While many of our techniques are applicable to a wide range of problems, a fundamental goal of this work is to demonstrate that applying QLS algorithms to real-world problems requires not only deep expertise in quantum algorithms but also significant domain-specific knowledge.
On the one hand, domain experts with limited exposure to quantum computing may overestimate how easily QLS algorithms can be applied to their problems, unaware of the inherent quantum limitations.
On the other hand, quantum computing specialists may be overly pessimistic, underestimating how small adjustments to real-world problems can bypass these quantum-specific challenges while still delivering meaningful results.
These domain-specific features not only influence the design and implementation of the relevant quantum algorithms, but also play a critical role in determining their feasibility and potential to outperform classical methods.
Close collaboration between domain and quantum experts is therefore essential to accurately assess the true potential and practicality of QLS algorithms for solving real-world problems.

Several prior works have studied quantum algorithms for solving linear systems, particularly in the context of preconditioning and discretized PDEs. Tong et al.~\cite{tong2021fast} introduced a framework for improving the condition‐number dependence of quantum linear‐system solvers from $O(\kappa^2)$ to $O(\kappa)$, alongside techniques for Green’s‐function computation and matrix‐function evaluation. While this was an important conceptual advance, subsequent work has established algorithms that achieve optimal $O(\kappa\log(1/\varepsilon))$ scaling under various assumptions. In particular, Costa et al.~\cite{costa2021optimal} proved such optimal scaling based on a discrete adiabatic theorem, which applies when the solution norm $\|x\|$ is known or can be efficiently estimated. Later numerical analysis by Costa et al.~\cite{Costa2025discreteadiabatic} showed that in practical settings the effective constant factor in that complexity bound is over $1200\times$ smaller than the analytic upper bound reported in~\cite{costa2021optimal}. Additionally, Low and Su~\cite{lowsu24qls} created a solver that reduces the number of oracle calls for state preparation, but in our case (since we assume the cost of state preparation is small) this advance is not super helpful. Dalzell~\cite{dalzell2024} further simplified the discrete‐adiabatic approach and provided explicit constants in the query complexity, assuming the solution norm is available. We adopt Dalzell’s formulation as the basis for our qubit and gate‐count estimates in Sec.~\ref{sec:explicit_counts}.

Other recent efforts have explored more structured problem classes, achieving potential quantum speedups. For example, Deiml and Peterseim~\cite{deiml2024fem} propose a quantum framework for the finite element method that shows promising performance, though its asymptotic behavior in higher dimensions needs further investigation. Additionally, Orsucci and Dunjko~\cite{orsucci2021qls} show that under certain structural and oracle assumptions (such as access to a Cholesky-like factorization) one can reduce the condition number dependence from $O(\kappa)$ to $O(\sqrt{\kappa})$. Lapworth and Sunderhauf~\cite{LapworthSunderhauf2025} also analyzed a number of different preconditioning techniques for implementing SPAI, Toeplitz, and circulant preconditioners on a quantum computer. They show empirical results that separately block encoding the preconditioner and system matrix is ineffective, and found that none of their techniques reduced the asymptotic scaling of the effective condition number.  Our proof transforms this empirical result into a rigorous mathematical result.

\section{Discretized Heterogeneous Elliptic PDEs}
A foundational class of problems in science and engineering involves modeling steady-state diffusion processes in heterogeneous media. These are described by 3D heterogeneous elliptic partial differential equations (PDEs). A general form of this equation is:
\begin{equation}\label{eq:general-poisson}
  \nabla \cdot (\K(\mathbf{r}) \nabla h(\mathbf{r})) = f(\mathbf{r}),
\end{equation}
where $h(\mathbf{r})$ is the field of interest (e.g., temperature, electrical potential, or pressure), $\K(\mathbf{r})$ is a spatially varying coefficient representing a material property (e.g., thermal conductivity, electrical conductivity, or permeability), and $f(\mathbf{r})$ is a source term. This equation models a steady-state condition where the flux of a quantity into any region is balanced by the flux out, plus any sources or sinks within the region.

As with most PDEs, eq.~\ref{eq:general-poisson} is often discretized using methods like finite differences, turning it into a large, sparse system of linear equations, which we will denote 
\begin{equation}
	Gh=f.
\end{equation}
If the domain is discretized into $N$ cells, then $G$ is an $N \times N$ matrix that encodes the geometry of the domain and the material property $\K$, while $h$ and $f$ are vectors of length $N$ that encode the field values and source terms at each cell, respectively.

\begin{figure}[H]
\centering
\includegraphics[height=0.13\textheight]{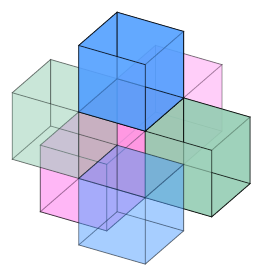} \hspace{2em}
\includegraphics[height=0.13\textheight]{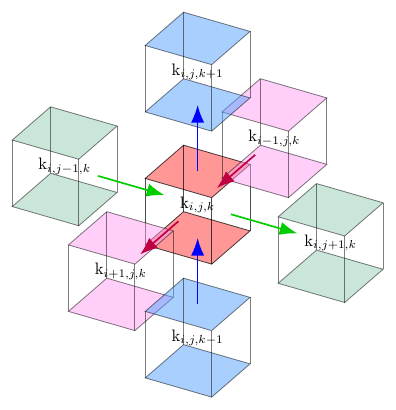}

	\caption{\label{fig:groundwater-flow-3d}Left: A 3D spatial grid representing a discretized domain of interest. Right: a visualization of a discretized heterogeneous elliptic PDE defined on the same 3D grid, where the cells are separated for ease of viewing. Each cell is given a spatially varying coefficient $\K_{i,j,k}$ (e.g., diffusivity, conductivity, or permeability) and a source term $f_{i,j,k}$. The goal is to determine the field $h_{i,j,k}$ that governs the PDE at each grid point.}
\end{figure}

For a concrete example, consider a domain discretized into a $N^{1/3} \times N^{1/3} \times N^{1/3}$ 3D grid of cubic cells. If the total domain has side length $L$, the cells will be of size $(\Delta x)^3 = L^3/N$. The flux from cell $(i,j, k-1)$ into an adjacent cell $(i,j,k)$ is proportional to the difference in the field $h$ between them and the material property $\K$ on the interface between them:
\begin{equation}
	(\K \nabla h)_{i,j, (k-1) \to k} = \K^{i,j, k-1}_{i,j,k}\frac{h_{i,j, k-1}-h_{i,j, k}}{\Delta x}.
\end{equation}
Here, $\K^{i,j, k-1}_{i,j,k}$ denotes the material property on the interface, which is often calculated as the geometric or harmonic mean of the property values in the adjacent cells. Summing the flows from all six neighbors for a central cell $(i,j,k)$ (as visualized in Fig. \ref{fig:groundwater-flow-3d}) yields the discrete version of eq.~\ref{eq:general-poisson}. The formula for $f_{i,j,k}$ is as follows: 
\begin{equation}\label{eq:pde-discrete-3d}
\begin{aligned}
& \frac{\K^{i-1,j,k}_{i,j,k}\bigl(h_{i-1,j,k}-h_{i,j,k}\bigr)}{(\Delta x)^2}
+ \frac{\K^{i,j-1,k}_{i,j,k}\bigl(h_{i,j-1,k}-h_{i,j,k}\bigr)}{(\Delta x)^2}
\\&+ \frac{\K^{i,j,k-1}_{i,j,k}\bigl(h_{i,j,k-1}-h_{i,j,k}\bigr)}{(\Delta x)^2}
+ \frac{\K^{i,j,k}_{i+1,j,k}\bigl(h_{i+1,j,k}-h_{i,j,k}\bigr)}{(\Delta x)^2}
\\&+ \frac{\K^{i,j,k}_{i,j+1,k}\bigl(h_{i,j+1,k}-h_{i,j,k}\bigr)}{(\Delta x)^2}
+ \frac{\K^{i,j,k}_{i,j,k+1}\bigl(h_{i,j,k+1}-h_{i,j,k}\bigr)}{(\Delta x)^2}.
\end{aligned}
\end{equation}
The three positive terms represent inflow from the back \((i-1)\), left \((j-1)\), and bottom \((k-1)\) neighbors, and the three negative terms represent outflow to the front \((i+1)\), right \((j+1)\), and top \((k+1)\) neighbors.

Following a row-by-row method for encoding the $N^{1/3} \times N^{1/3} \times N^{1/3}$ 3D grid into a vector of length $N$ gives
\begin{equation*}\label{eq:vec-encoding}
\begin{tikzpicture}[baseline=(current bounding box.center), scale=0.75]
		\node at (2, 2.6) {\textbf{Layer 1} ($i=0$)};
		
		\draw (0, 0) rectangle (4, 2);
		\draw (2, 0) -- (2, 2);
		\draw (0, 1) -- (4, 1);
		\node at (1, 1.5) {$f_{0,0,0}$};
		\node at (3, 1.5) {$f_{0,1,0}$};
		\node at (1, 0.5) {$f_{0,0,1}$};
		\node at (3, 0.5) {$f_{0,1,1}$};

		\begin{scope}[shift={(5,0)}]
			\node at (2, 2.6) {\textbf{Layer 2} ($i=1$)};
			
			\draw (0, 0) rectangle (4, 2);
			\draw (2, 0) -- (2, 2);
			\draw (0, 1) -- (4, 1);
			\node at (1, 1.5) {$f_{1,0,0}$};
			\node at (3, 1.5) {$f_{1,1,0}$};
			\node at (1, 0.5) {$f_{1,0,1}$};
			\node at (3, 0.5) {$f_{1,1,1}$};
		\end{scope}
	\end{tikzpicture}
	=
	\begin{pmatrix}
		f_{0,0,0} \\ f_{0,0,1} \\ f_{0,1,0} \\ f_{0,1,1} \\
		f_{1,0,0} \\ f_{1,0,1} \\ f_{1,1,0} \\ f_{1,1,1}
	\end{pmatrix}
\end{equation*}
This implies that $f_{i,j,k} = f_{k + j N^{1/3} + i N^{2/3}}$. If we define $a\equiv k + j N^{1/3} + i N^{2/3}$, then eq.~\ref{eq:pde-discrete-3d} can be rearranged such that we get the following formula for $f_a$:
\begin{equation*}
\begin{aligned}
&h_{a - N^{2/3}}\;\frac{\K^{i-1,j,k}_{i,j,k}}{(\Delta x)^2}
\;+\;
h_{a - N^{1/3}}\;\frac{\K^{i,j-1,k}_{i,j,k}}{(\Delta x)^2}
\;+\;
h_{a - 1}\;\frac{\K^{i,j,k-1}_{i,j,k}}{(\Delta x)^2}\\
&
+\;h_a\;\Biggl(
  -\frac{\K^{i-1,j,k}_{i,j,k} + \K^{i,j,k}_{i+1,j,k}}{(\Delta x)^2}
  -\frac{\K^{i,j-1,k}_{i,j,k} + \K^{i,j,k}_{i,j+1,k}}{(\Delta x)^2}
  \\& \qquad \qquad \quad-\frac{\K^{i,j,k-1}_{i,j,k} + \K^{i,j,k}_{i,j,k+1}}{(\Delta x)^2}
\Biggr)\\
&
+\;h_{a + 1}\;\frac{\K^{i,j,k}_{i,j,k+1}}{(\Delta x)^2}
\;+\;
h_{a + N^{1/3}}\;\frac{\K^{i,j,k}_{i,j+1,k}}{(\Delta x)^2}
\;+\;
h_{a + N^{2/3}}\;\frac{\K^{i,j,k}_{i+1,j,k}}{(\Delta x)^2}.
\end{aligned}
\end{equation*}
Combining the linear equations of this form from each cell produces a system of equations defined by the matrix $G$. $G$ is given by the following formula, where we removed a multiplicative factor of $(\Delta x)^{-2}$ from each condition:
\begin{equation}\label{eq:G-def}
\begin{aligned}
G_{a,b} =
\begin{cases}
-\K^{i-1,j,k}_{i,j,k}, & b = a - N^{2/3}, \\[4pt]
-\K^{i,j-1,k}_{i,j,k}, & b = a - N^{1/3}, \\[4pt]
-\K^{i,j,k-1}_{i,j,k}, & b = a - 1, \\[4pt]
\big(\K^{i-1,j,k}_{i,j,k} + \K^{i,j-1,k}_{i,j,k} + \K^{i,j,k-1}_{i,j,k} 
   \\[3pt] \ +\K^{i,j,k}_{i+1,j,k} + \K^{i,j,k}_{i,j+1,k} + \K^{i,j,k}_{i,j,k+1} \big), & b = a, \\[4pt]
-\K^{i,j,k}_{i,j,k+1}, & b = a + 1, \\[4pt]
-\K^{i,j,k}_{i,j+1,k}, & b = a + N^{1/3}, \\[4pt]
-\K^{i,j,k}_{i+1,j,k}, & b = a + N^{2/3}, \\[2pt]
0, & \text{otherwise.}
\end{cases}
\end{aligned}
\end{equation}
where we have added an overall negative sign to make $G$ positive definite.
Note that this formula is only valid for $1<i,j,k<N^{1/3}$, for terms on the boundary of the grid additional elements of $G_{a,b}$ will also be 0.
In many physical systems, Dirichlet boundary conditions are imposed, which involves setting certain rows of $G$ to be 1 on the diagonal and 0 elsewhere, with corresponding adjustments to $f$. The resulting matrix $G$ is sparse, Hermitian, and positive definite. Classically, systems of this type are often solved using iterative methods like the Conjugate Gradient (CG) method. The CG method scales with a runtime of $O(N\sqrt{K(G)} \log (1/\varepsilon))$, where $K(G)$ is the condition number of $G$. In practice, however, the convergence rate of the CG algorithm with appropriate preconditioning is $O(N)$ or $O(N \log N \cdot \log(1/\epsilon))$~\cite{strang2007computational}.

Even with these advanced classical techniques, state-of-the-art methods struggle to model real world domains with more than three or four orders of magnitudes of resolution~\cite{hyman2015dfnWorks}.
This is generally because of memory rather than speed constraints.
Coarse-graining is often effective at skirting such memory demands by not considering all length scales at once.
However, this approach fails for problems where there are complex interactions the span many different length scales.
A key motivating example is modeling the flow of underground fluids (e.g., water, oil) through porous and fractured media. In this case, the governing PDE is known as the groundwater flow equation, where $h$ is the hydraulic head (a measure of fluid pressure), $\K$ is the permeability of the medium, and $f$ represents fluid sources or sinks.
For these problems, critical physical properties like percolation depend on simultaneous interactions across fractures of all length scales~\cite{Golden_2022}.
Capturing this behavior can require modeling kilometer-sized domains with fractures down to 1 cm in scale.
The resulting matrix would contain $O(10^{15})$ cells (and non-zero entries) in 3D, requiring $O(100)$ petabytes of memory, beyond the limits of modern computing systems.
Therefore, accurate modeling of these systems requires fine-scale resolution beyond the reaches of classical high-performance computing.

\section{Solving Discretized Poisson PDEs with QLS} \label{sec:QLS_steps_for_paper}
The first QLS algorithm, by Harrow, Hassidim, and Lloyd (HHL)~\cite{harrow2009quantum}, solves the system $Ax=b$ by preparing a state $\ket{x} = \ket{A^{-1}b}$ to accuracy $\epsilon$ in time $O(\kappa^2 \log N/\epsilon)$ where $A\in \mathbb{R}^{N\times N}$ (generally $A\in \mathbb{C}^{N\times N}$, but throughout this paper we are only considering real system matrices) and has effective condition number $\kappa$.
While this provides a theoretical exponential improvement over naive classical methods (which scale as $O(N^3)$), the algorithm relies heavily on quantum phase estimation and amplitude amplification techniques, making it complex to implement in practice.

More recent advancements have improved both the complexity and practicality of QLS algorithms.
Notably, Costa et al. \cite{costa2021optimal} introduced a quantum algorithm that builds on qubitization methods and employs a novel discrete adiabatic theorem, achieving a complexity of $O(\kappa \log(1/\epsilon))$ and saturating known lower bounds in both $\kappa$ and $\epsilon$.
This method avoids the need for variable-time amplitude amplification or truncated Dyson series, which were common in previous methods, making it more efficient and easier to implement. More recently, Dalzell \cite{dalzell2024} created a QLS algorithm that eliminates the need for ansatz state preparation, as used in Costa et al. \cite{costa2021optimal}. This makes it even easier to analyze/implement and also saturates the known lower bound, now with improved constant factors.
Despite these advances, several significant hurdles must be cleared before this (or any future) QLS algorithm can be employed to actually solve a linear system. 

\paragraph{Efficient State Preparation of $\ket{b}$.} The QLS algorithm requires the source vector $b$ to be prepared as a quantum state, $\ket{b}$. This preparation must be efficient, meaning its cost should not dominate the overall runtime. For many physical problems, including our groundwater flow example, the source terms $f$ are often sparse, representing localized inputs like injection or extraction wells, and can be encoded efficiently~\cite{henderson2023addressingquantumsfineprint}.

\paragraph{Efficient Block Encoding of the Matrix $A$.} State-of-the-art QLS algorithms require a block encoding $U_A$, which is often highly nontrivial. Fortunately, matrices arising from the discretization of local PDEs via finite difference methods are highly structured and sparse, which often permits an efficient block encoding, as we will detail in Sec. \ref{sec:encode-G}.

\paragraph{The Condition Number $\kappa$ Must Be Manageable.} The runtime of all known QLS algorithms depends at least linearly on the effective condition number, $\kappa$. For 3D Poisson-type problems, $\kappa$ scales polynomially with the number of grid cells $N$ (typically as $O(N^{2/3})$). This scaling can severely degrade the quantum speedup, potentially negating any exponential advantage over classical methods. While classical solvers rely on preconditioning to reduce $\kappa$, applying preconditioners in a quantum context is a significant challenge, as we will discuss in Sec. \ref{sec:preconditioning}.

\paragraph{Efficient Extraction of a Useful Answer.} A QLS algorithm produces a quantum state $\ket{x}$ proportional to the solution vector $x$. Reconstructing the full classical vector $x$ from this state would require a number of measurements that scales with $N$, eliminating any quantum advantage. Therefore, the desired solution must be a specific, efficiently extractable property of $\ket{x}$. In physical problems, we are often interested in local observables or global properties, which can be estimated efficiently from $\ket{x}$ using a small number of measurements. This will be explored further in Sec.~\ref{sec:extract-x}.

The following sections address these requirements in detail, using the groundwater flow problem as a concrete test case to assess the viability and remaining challenges of a quantum solution.

\section{Block Encoding} \label{sec:encode-G}
In this section we present a general block-encoding strategy for the discrete 3D heterogeneous Poisson system, extending the structured-oracle framework of~\cite{Snderhauf2024} from the 2D Laplacian to fully three-dimensional problems. Given any coefficient field that can be specified by an arithmetic rule or lookup table with $D=O(\mathrm{polylog}\,N)$ distinct values, the block encoding has cost $O(\mathrm{polylog}\,N)$. This structural assumption plausibly holds for a wide class of diffusion, conductivity and elasticity models, and we discuss its implementation in detail for the groundwater flow problem.

A block encoding of a matrix $A$ is a unitary matrix $U_A$ that contains $A/\alpha$ as a sub-matrix in its upper-left corner: 
\begin{equation} \label{eq:block_encoding_structure}
	U_A = \begin{pmatrix} A/\alpha & * \\ * & * \end{pmatrix},
\end{equation}
where $\alpha \ge \|A\|$ (throughout this paper we use $\|\cdot \|$ to denote the spectral norm) is required to ensure that the scaled matrix $A/\alpha$ has spectral norm at most one. 
The asterisks $*$ represent arbitrary entries (of arbitrary sizes) needed to satisfy the unitarity of $U_A$.

To construct a block encoding of the discretized Poisson operator $G$, we first rescale $G\to G'$ where $\|G'\| \le 1$.
Applying the Gershgorin circle theorem, we can bound the spectral norm by
\begin{equation} \label{eq:||G||_bound}
	\|G\| \le 12\K_{\text{max}}/(\Delta x)^{2} = 12k_{\text{max}} \cdot \frac{N^{2/3}}{L^2},
\end{equation}
where $\K_{\text{max}}$ is the largest element of $\K$.
This value comes from a row in $G$ corresponding to a cell with $k_{\text{max}}$ surrounded by other cells with $\K_{\text{max}}$.
For most physical systems governed by elliptic PDEs, $\K_{\text{max}}$ is a constant that depends on the intrinsic material property of the domain (diffusivity, permeability, conductivity, etc.), and is independent of how the PDE is discretized -- that is, independent of the mesh resolution (i.e., the total number of grid cells $N$).

In the the groundwater flow problem, for example, fracture permeability is often modeled as proportional to the length of the fracture to some power $\beta$, where $\beta$ depends on the type of fractured material~\cite{hyman2016fracture}.
We assume that the largest fracture in $R$ is contained entirely within $R$, that is, it has length $\le L$, and so $\K_{\text{max}} \le L^{\beta}$. More generally, the size of the largest fracture will be limited (e.g., by the size of the earth), making $\K_{\text{max}}$ a constant.

We can then define a new constant $a_{\text{max}} = 12L^{-2}\K_{\text{max}}$ and normalize $G$ as $G' = G/(a_{\text{max}} \cdot N^{2/3})$. This gives,
\begin{equation} \label{eq:bound_on_G'_norm}
    \|G'\| = \|G/(a_{\text{max}} \cdot N^{2/3})\| \leq 1
\end{equation}
as required.

Structurally, $G$ closely resembles the discrete Laplacian operator.
In particular, eq.~\ref{eq:G-def} reduces to the Laplacian $\Delta$ in the case of constant $\K$, where we have (modulo constant factors)
\begin{equation}\label{eq:def-laplacian}
	\Delta_{a,b} = \begin{cases}
		-1, &b=a\pm N^{2/3}, b=a \pm N^{1/3}, b=a\pm 1,\\
		6, &b = a,\\
		0  &\text{else.}
	\end{cases}
\end{equation}
Recently, an exact block encoding of the 2D Laplacian was introduced by Sunderhauf et al.~\cite{Snderhauf2024}, which with some modifications, can be used to block encode the 3D Laplacian. We will not explicitly do it in this paper as it is a special case of the matrix $G'$, and thus follows from our block encoding for $G'$ in App. \ref{sec:BE_GW_flow}.
Sunderhauf's algorithm is designed for matrices $A$ which contain structured data with relatively few distinct non-zero elements, e.g. Toeplitz or binary tree matrices.
The primary components of this block encoding are a data loading oracle, which encodes matrix values, and structure oracles, which specify the positions of each nonzero entry.
The data loading oracle has the form:
\begin{equation}
	O_{\text{data}} = \sum_{d=0}^{D-1} R_X\left( 2 \arccos G'_d \right) \otimes \ket{d}\bra{d},
\end{equation}
where $D$ is the total number of distinct entries in $G'$.
This data is structured using transposition and column oracles $O_t$ and $O_c$, which are defined as: 
\begin{equation}
	O_t\ket{d}\ket{m} = \ket{d}\ket{m'}, \quad O_c\ket{d}\ket{m} = \ket{j}\ket{s_c}. 
\end{equation}
The transposition oracle allows us to map the $m$-th occurrence of the $d$-th distinct value to its symmetric counterpart $m'$, related to $m$ by transposition. Additionally, the column oracle 
returns the column index $j$ and sparsity label $s_c$ that corresponds to the $m$-th appearance of the $d$-th distinct element.
Since the 3D heterogeneous Poisson system matrix is sufficiently structured, such oracles can be efficiently implemented on a quantum computer.
In App. \ref{sec:BE_GW_flow}, we provide an explicit construction of $G'$ and prove it has gate complexity $O(\log N + D\log D)$. Notice that this, nor our final QLS runtime, has any dependence on the matrix sparsity. In App.~\ref{sec:BE_GW_flow}, we show that the maximum sparsity of any row/column is bounded by $7$, so this reliance on sparsity is removed when we are considering asymptotic notation. 

To bound $D$, recall that the non-zero terms in eq. \ref{eq:G-def} only depend on the $\K$ values for adjacent cells. Suppose the domain is discretized such that there are at most $F$ distinct values that $\K$ can take across all cells. Then, using geometric or harmonic means, the number of possible interface values (i.e., the values assigned to the permeability at each face between neighboring cells) is at most $F^2$, corresponding to all possible pairwise combinations of adjacent cell coefficients.

Since the discrete 3D Poisson matrix includes up to 6 off-diagonal entries per row (corresponding to the 6 face-adjacent neighbors), and each such term is defined by one of the $F^2$ interface values, the total number of distinct patterns of nonzero entries per row is upper-bounded by the number of 6-tuples over $F^2$ values. Therefore, the number of distinct nonzero values in $G$ satisfies
\begin{equation} \label{eq:distinct_elements_vs_F}
    D = O\left( (F^2)^6 \right) = O(F^{12}).
\end{equation}
This provides an upper bound on the number of unique matrix entries that need to be block encoded.

As an example, in App. \ref{sec:D_bound_gw_flow} we show that for groundwater flow that $F$ scales as $\text{polylog}(N)$, for fracture patterns with arithmetic descriptions. As an example of typical values for $F$, (Greer et al `22) found that in a real world study from the Topopah Spring Aquifer in Nevada, fracture sizes range from $1.5$ to $125$ m in a $250$ m by $250$ m by $100$ m domain.

\section{Effective Condition Number} \label{sec:condition_number}
State-of-the-art QLS algorithms~\cite{costa2021optimal, dalzell2024} require the system matrix $A$ to be rescaled to a matrix $A' = A/\alpha_A$ such that $\alpha_A \geq \|A\|$. The effective condition number is then $\kappa = \|(A')^{-1}\| = \alpha_A \cdot \|A^{-1}\| \geq K(A)$, where $K(A) = \|A\| \cdot \|A^{-1}\|$ is the condition number of $A$. In the next subsection, we examine the implications of this rescaling.

\begin{figure}[H]
    \centering    \includegraphics[width=\columnwidth]{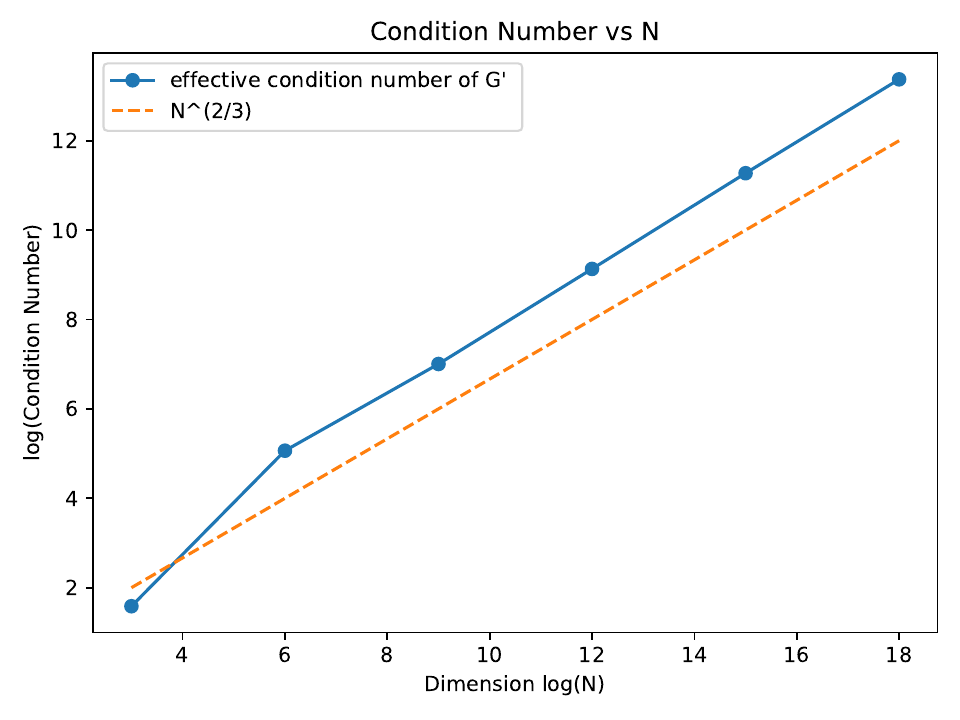}
    \caption{This figure shows the empirical scaling of the condition number (equivalent to the effective condition number of the system prior to preconditioning) for solving the groundwater flow equation. The simulation uses the 3D pitchfork fracture network described in Fig.~\ref{fig:3D_pitchfork}. As the dimension $N$ increases, the condition number scales as $\kappa = O(N^{2/3})$, consistent with the theoretical analysis in the beginning of Sec.~\ref{sec:condition_number}.}
	\label{fig:cond_num_scaling}
\end{figure}

In Sec. \ref{sec:encode-G} we showed that $G' = \frac{G}{O(N^{2/3})}$, so the effective condition number is $$\kappa = \|(G')^{-1}\| = O(N^{2/3}) \cdot \|G^{-1}\|.$$ 
To upper bound $\|G^{-1}\|$, we can use the discrete Poincaré inequality to get $\|G^{-1}\| = O(1)$, as seen in App. \ref{sec:Proof of ||G^{-1}||}. This gives us that $\kappa$ scales as $O(N^{2/3})$. An example of this scaling for groundwater flow  can be seen in Fig. \ref{fig:cond_num_scaling}. Combined with the other findings in this paper, we get the quantum runtime is $O(N^{2/3} \text{ polylog }N \cdot \log(1/\epsilon))$, which is a speed up over classical approaches which scale as $O(N \log N \cdot \log(1/\epsilon))$. The hope is to further reduce $\kappa$ (and thus reduce the complexity of running QLS algorithms), but this is challenging. The next subsection explains some challenges with reducing the effective condition number.

\subsection{Reducing the effective condition number} \label{sec:preconditioning}
While current QLS algorithms theoretically achieve optimal scaling with respect to the effective condition number $\kappa$, practical applications hinge on our ability to reduce $\kappa$ efficiently.
In this section, we show that separately block encoding preconditioners and their system matrix, even with singular value amplification, does not work to achieve asymptotic quantum speed ups. As a result, more advanced techniques are needed to further improve the scaling of QLS algorithms for solving 3D heterogeneous Poisson equations.

To start, preconditioning is a well-established technique in the context of classical linear system solvers, where the primary objective is to reduce the condition number of the system, thereby improving the convergence rate of iterative methods.
When the condition number of the system matrix $A$  scales poorly with the problem size $N$, a preconditioner matrix $M$ can be introduced such that the condition number of the product $M A$ exhibits more favorable scaling with $N$.
Instead of directly solving the original system $A x = b$, the preconditioned system $MAx = Mb$ is solved, where the improved condition number leads to faster convergence. 

Block encoding techniques provide a natural and efficient way to implement preconditioners for QLS.
In the block encoding framework, given block encodings $U_M$ for the preconditioner $M$ and $U_A$ for the system matrix $A$, their product $U_M U_A$ directly yields a block encoding of the matrix product $MA$. As seen in eq. \ref{eq:block_encoding_structure}, this product will result in a normalization factor of $\alpha_A \cdot \alpha_M$.
The total cost of constructing this block encoding is then simply the sum of the costs associated with $U_M$ and $U_A$.

As an example, in a previous work~\cite{Golden_2022}, we argued that in two dimensions the inverse Laplacian $\Delta^{-1}$ is an effective preconditioner for the groundwater flow matrix $G$, and that it reduces the scaling of the condition number from $O(N)$ to approximately $O(\sqrt{N})$ for multiscale fractal fracture networks. We show in App. \ref{sec:BE_inv_laplacian} that $\Delta^{-1}$ can be implemented in $O(\log N)$ time.

Despite having good preconditioners that are easy to block encode, block encoding $A$ and $M$ separately does not enable QLS to benefit from the improved condition number. The effective condition number of the system resulting from separately block encoding the preconditioner and system matrix (i.e. $U_M U_A = U_{MA})$ is $\kappa = \alpha_M \cdot \alpha_A \cdot \|A^{-1} M^{-1}\|$, and QLS does not benefit from the smaller condition number, $K(MA)$.

To start, recall that the spectral norm is submultiplicative. This means that for any matrices $B$ and $C$, $\|BC\| \leq \|B\| \cdot \|C\|$. Using this, we get that for any invertible matrices $B$ and $C$ that
\begin{equation}\label{eq:submult_2norm_bound}
    \|B^{-1}C^{-1}\| \cdot \|C\| \geq \|B^{-1}\| \iff \|B^{-1}C^{-1}\| \geq \frac{\|B^{-1}\|}{\|C\|}
\end{equation}
Now, combining this bound along with the fact that $\alpha_M \geq \|M\|$ and $\alpha_A \geq \|A\|$, we get a lower bound on $\kappa$:
\begin{align*}
    \kappa &= \alpha_M \cdot \alpha_A \cdot \|A^{-1} M^{-1}\|\\
    &\geq \|M\| \cdot \alpha_A \cdot \|A^{-1} M^{-1}\|\\
    &\geq \|M\| \cdot \alpha_A \cdot \frac{\|A^{-1}\|}{\|M\|}\\
    &= \alpha_A \cdot \|A^{-1}\|\\
    &= \kappa(A)
\end{align*}
So for any QLS algorithm, if you block encode the system matrix and the preconditioner separately, then multiply them together to get a preconditioned matrix, you can never reduce the effective condition number. So using this approach, there is no way to achieve a quicker QLS runtime.

Applying SVA (singular value amplification) to fix this issue also fails to improve the asymptotic quantum runtime. SVA would improve the block encoding normalization factors and transform the embedding of the matrix from $\frac{MA}{\alpha_A \cdot \alpha_M} \rightarrow \frac{MA}{\|MA\|}$. As seen in Lapworth and Sunderhauf \cite{LapworthSunderhauf2025}, the additional complexity this process incurs outweighs the benefits. You might be able to get constant time speed ups, but no asymptotic changes in the complexity. They also show a way of reducing the subnormalization factor of the composed block encoding $U_{MA}$ from $\alpha_A \alpha_M $ to $ \frac{\alpha_A \alpha_M}{\gamma_A \gamma_M}$, where we can choose $\gamma_A$ and $\gamma_M$. This does not improve the scaling as it requires individually decreasing the subnoralization factor of $U_A$ from $\alpha_A$ to $\alpha_A/\gamma_A$ and the subnoralization factor of $U_M$ from $\alpha_M$ to $\alpha_M/\gamma_M$. Due to block encoding restrictions, $\alpha_A/\gamma_A \geq \|A\|$ and $\alpha_M/\gamma_M \geq \|M\|$, so the best subnormalization factor after this process is $\frac{\alpha_A \alpha_M}{\gamma_A \gamma_M} = \|A\| \cdot \|M\|$. Therefore, applying SVA will not improve the overall scaling of QLS algorithms.

This demonstrates the ineffectiveness of block encoding preconditioners separately from the matrix $A$. Despite having good preconditioning matrices, by simply block encoding them we can achieve no scaling improvement. Lapworth and Sunderhauf \cite{LapworthSunderhauf2025} also tried a number of other preconditioning methods that would be applicable for discretized Poisson PDEs, but again they only found constant time improvements in the effective condition number.  

Due to the structure of the system matrix in 3D, we are able to achieve a quantum speedup; however, for 2D discretized Poisson PDEs, $K(G)$ scales as $O(N)$. This results in a quantum runtime of $O(N \text{ polylog }N \cdot \log(1/\epsilon))$, performing worse than classical algorithms that scale as $O(N \log N \cdot \log(1/\epsilon))$. For discretized Poisson PDEs, if we want to improve our quantum speed up in 3D, or achieve a speed up in 2D, we must find better preconditioning techniques.

Several recent works offer promising ideas that may help improve the asymptotic scaling of the effective condition number for our QLS algorithm. Deiml and Peterseim~\cite{deiml2024fem} introduce a quantum realization of the finite element method that is promising. They importantly do a combined preconditioning approach that does not conflict with our separate preconditioning bound. In Theorem 6.5 of their paper, they achieve a runtime of $O(\frac{1}{\varepsilon} \text{ polylog } \varepsilon)$, ignoring block encoding costs. They also make the same assumption about an efficient look-up table for the values in the system matrix. With a standard block encoding cost, like the one seen in App.~\ref{sec:BE_GW_flow}, their algorithm would achieve a total runtime of $O(\frac{\text{polylog }N}{\varepsilon} \text{ polylog } \varepsilon)$. In App.~\ref{sec:scaling_in_eps}, however, we show that in practical examples $1/\varepsilon$ can scale as $O(N)$. This would result in a worse asymptotic bound than what is seen in our algorithm, but highlights that joint preconditioning is possible in certain scenarios. Future works are needed to further explore the possibility of these ideas though. Separately, Orsucci and Dunjko \cite{orsucci2021qls} develop techniques for solving classes of positive-definite quantum linear systems, and most notably show that if the system matrix admits an efficiently implementable decomposition $G = LL^\dagger$, akin to the classical Cholesky decomposition, then you can achieve an effective condition number of $\sqrt{\kappa(G)}$. By efficiently implementable they mean that $G$ can be expressed as a sum of local, positive-definite Hamiltonians, each acting non-trivially on only $\text{polylog}(N)$ qubits. These results suggest that preconditioning techniques that exploit problem-specific structure (such as that arising from discretized 3D heterogeneous Poisson equations) may enable further QLS speedups.

\section{Information Extraction}\label{sec:extract-x}
As discussed in~\cite{henderson2023addressingquantumsfineprint}, the Hadamard test is a straightforward way to determine the average value of $x$ in specific regions of the domain (e.g., corresponding to a well screen where a physical measurement could be taken in the fracture flow context). In a real-world case the observable of interest is coarse-grained, e.g. for groundwater flow the average pressure in a cylindrical subregion of $R$ with diameter typically on the order of 10cm and height on the order of 1m  (i.e., the well diameter and the screen height).
That means that refining the grid past 0.1m resolution will not significantly affect the necessary precision for information extraction.
Furthermore, it is straightforward to prepare a state $\ket{\phi}$ which captures the same physical subregion regardless of grid refinement.
To show this, we take $N = 2^{3\ell}$, and let $n = N^{1/3} = 2^\ell$, and assume we are interested in the physical region corresponding to the single cell at entry $(i,j,k)$.
This corresponds to the $r$th element of $x$, where $r = k+nj+n^2i$, and means we want to measure
\begin{equation}
	\braket{\phi^{(n)}|x^{(n)}}
\end{equation}
where $\ket{\phi^{(n)}} = \ket{r}$ is the $3\ell$-bit binary representation of the integer $r$, and $x^{(n)}$ is the solution to groundwater flow equation with an $n\times n \times n$ grid.

Now let us consider what happens when we double the grid resolution and want to measure the total pressure in the same physical area.
Since we have doubled the resolution, the area of interest is now represented by eight cells:
\begin{equation}
\begin{tikzpicture}[scale=0.8, baseline={(current bounding box.south)}]

  \draw[thick] (0,0,0) -- (2,0,0) -- (2,2,0) -- (0,2,0) -- cycle;
  \draw[thick] (0,0,2) -- (2,0,2) -- (2,2,2) -- (0,2,2) -- cycle;
  \draw[thick] (0,0,0) -- (0,0,2);
  \draw[thick] (2,0,0) -- (2,0,2);
  \draw[thick] (2,2,0) -- (2,2,2);
  \draw[thick] (0,2,0) -- (0,2,2);

  \foreach \x in {1} {
    \draw[dashed] (\x,0,0) -- (\x,2,0);
    \draw[dashed] (\x,0,2) -- (\x,2,2);
    \draw[dashed] (\x,0,0) -- (\x,0,2);
    \draw[dashed] (\x,2,0) -- (\x,2,2);
  }
  \foreach \y in {1} {
    \draw[dashed] (0,\y,0) -- (2,\y,0);
    \draw[dashed] (0,\y,2) -- (2,\y,2);
    \draw[dashed] (0,\y,0) -- (0,\y,2);
    \draw[dashed] (2,\y,0) -- (2,\y,2);
  }
  \foreach \z in {1} {
    \draw[dashed] (0,0,\z) -- (2,0,\z);
    \draw[dashed] (2,0,\z) -- (2,2,\z);
    \draw[dashed] (0,2,\z) -- (2,2,\z);
    \draw[dashed] (0,0,\z) -- (0,2,\z);
  }

  \fill[red, opacity=0.4] (1,1,1) -- (2,1,1) -- (2,2,1) -- (1,2,1) -- cycle;
  \fill[red, opacity=0.4] (1,1,2) -- (2,1,2) -- (2,2,2) -- (1,2,2) -- cycle;
  \fill[red, opacity=0.4] (1,1,1) -- (1,1,2) -- (2,1,2) -- (2,1,1) -- cycle;

  \node[below=1cm of {(0.5,0,0)}, align=center] {Original Cell};
\end{tikzpicture}
\hspace{1cm}
\hspace{1cm}
\begin{tikzpicture}[scale=0.5, baseline={(current bounding box.south)}]
  \foreach \x in {0,1,2,3,4} {
    \foreach \y in {0,1,2,3,4} {
      \draw[gray, thin] (\x,0,\y) -- (\x,4,\y);
      \draw[gray, thin] (0,\x,\y) -- (4,\x,\y);
      \draw[gray, thin] (\x,\y,0) -- (\x,\y,4);
    }
  }

\foreach \x in {2,3} {
  \foreach \y in {2,3} {
    \foreach \z in {2,3} {
      \fill[red, opacity=0.4]
        (\x,\y,\z) -- ++(1,0,0) -- ++(0,1,0) -- ++(-1,0,0) -- cycle; 
      \fill[red, opacity=0.4]
        (\x,\y,\z) -- ++(1,0,0) -- ++(0,0,1) -- ++(-1,0,0) -- cycle; 
      \fill[red, opacity=0.3]
        (\x,\y,\z) -- ++(0,1,0) -- ++(0,0,1) -- ++(0,-1,0) -- cycle; 
    }
  }
}
  \node[below=1cm of {(1.25,0,0)}, align=center] {Refined Grid};
\end{tikzpicture}
\end{equation}

where the areas shaded in red represent the physical region of interest.
In the refined mesh, the row and columns of these shaded cells are given by
\begin{align*}
(i,j,k) \rightarrow \{&(2i, 2j, 2k), (2i, 2j, 2k+1), (2i, 2j+1, 2k), \\& (2i, 2j+1, 2k+1), (2i+1, 2j, 2k), (2i+1, 2j, 2k+1),\\& (2i+1, 2j+1, 2k), (2i+1, 2j+1, 2k+1)\}
\end{align*}
which corresponds to the indices 
\begin{equation}
\begin{aligned}
    \{r'+4n^2+2n+1, r'+4n^2+2n, r'+4n^2+1,\\ r'+4n^2, r'+2n+1, r'+2n, r'+1, r' \}
\end{aligned}
\end{equation}
for $r' = 2k + 2n (2j + 2n \cdot 2i)$.
This means that for the refined system $G^{(2n)} x^{(2n)} = f^{(2n)}$, we want to measure
\begin{equation}
	\braket{\phi^{(2n)}|x^{(2n)}}
\end{equation}
where
\begin{align*}
	\ket{\phi^{(2n)}} = \frac{1}{\sqrt{8}}(&\ket{r'} + \ket{r'+1} + \ket{r'+2n} + \ket{r'+2n+1} \\ &+\ket{r'+4n^2} + \ket{r'+4n^2+1} \\&+\ket{r'+4n^2+2n} + \ket{r'+4n^2+2n+1}).
\end{align*}
In the binary representation of $r$ with $3\ell$ qubits/digits, the first $\ell$ digits represent $i$, the next $\ell$ digits represent $j$, and the last $\ell$ digits represent $k$.
Then, in the binary representation of $r' = 2k + 2n (2j + 2n \cdot 2i)= 2k + 2^{\ell+2}j + 2^{2\ell + 3}i$ with $3\ell+3$ qubits, the first $\ell+1$ digits will be the binary representation of $k$ shifted to the left by one, the next $\ell+1$ digits represent $j$ shifted to the left by one, and similarly, the last $\ell+1$ digits represent $k$ shifted one to the left.
In other words, if the binary representation of $r$ has 1's at locations $\{l_a\}$, then $r'$ will have 1's at locations
\begin{equation}
	l'_a = \begin{cases}
		l_a + 1,\quad a < \ell\\
		l_a + 2,\quad \ell\leq a <2\ell \\
        l_a+3,\quad a\geq2\ell
	\end{cases}
\end{equation}
Therefore $X$ gates on the qubits $\{l'_a\}$ will prepare the state $\ket{r'}$.
Note that for a number $s$ whose binary representation in the $i$-th digit is zero,
\begin{equation}
	H^{(i)}\ket{s} = \frac{1}{\sqrt{2}}\left(\ket{s} + \ket{s+2^i}\right),
\end{equation}
where $H^{(i)}$ means applying a Hadamard to the $i$-th qubit.
Since the binary representation of $r'$ has zeros at the $0$-th, $(\ell+1)$-th, and $(2\ell+2)$-th digits, we have
\begin{equation}
	\ket{\phi^{(2n)}} = H^{(2\ell+2)}H^{(\ell+1)}H^{(0)}\ket{r'}.
\end{equation}
Therefore doubling the grid resolution requires adding only three qubits and three $H$ gates for the information extraction.
\section{Explicit Gate and Qubit Count}
\label{sec:explicit_counts}

In this section we provide an explicit estimate of the gate and qubit resources required to apply the adiabatic-inspired linear system algorithm of Dalzell~\cite{dalzell2024} to the block encoding of $G'$ constructed in App~\ref{sec:BE_GW_flow}.

\subsection{Mapping of the Block Encoding to Dalzell’s Complexity Model}
In this analysis we will focus on Dalzell’s QLS algorithm~\cite{dalzell2024} as it exhibits better constant factors than that in Costa et al.~\cite{costa2021optimal}. First, let $G_{total}$ represent the total number of elementary gates used in Danzell's algorithm. Theorem~4 of his paper shows
\begin{equation}
    Q \le 56.0\kappa + 1.05\kappa \ln\left(\frac{\sqrt{1-\varepsilon^2}}{\varepsilon} \right)
    + 2.78 \ln (\kappa)^3 + 3.17,
    \label{eq:dalzell-Q}
\end{equation}
and the QLS algorithm makes an expected $Q$ total queries to $U_{G'}$, $U_{G'}^\dagger$ and their controlled versions. It also makes $2Q$ queries to $U_b$, $U_b^\dagger$, and their controlled versions (where $U_b$ is the unitary that generates the state $\ket{b}$.\\
The controlled versions of these unitaries are going to be computationally more expensive than the non-controlled versions. So, for an upper bound we can assume we only makes calls to the controlled versions. Additionally, we can bound this further by noticing that to go from the normal to controlled version we just need to convert every gate to its controlled version. From Lubinski et al.~\cite{Lubinski23}, the most computationally intensive gate to do this with is converting a CNOT gate to a toffoli gate, which adds a multiplicative factor of 16. So, we can write $G_{total}$ as
\begin{equation}
    G_{total} \leq 16Q \cdot g_{U_{G'}} + 32Q\cdot g_{U_b}, 
    \label{eq:Gtotal}
\end{equation}
where $g_{U_{G'}}$ and $g_{U_b}$ denote the gate counts for a single query of $U_{G'}$ and $U_b$ respectively.

\subsection{Exact Gate Count for the Block Encoding of $G'$}

From the explicit circuit in Appendix E, the block-encoding $U_{G'}$ uses the following registers:
\begin{align*}
    d^{ind} &: 2~\text{qubits},\\
    d^{val} &: \lceil \log_2 D' \rceil \text{ qubits},\\
    m^{hi}  &: 1~\text{qubit},\\
    m^{lo}  &: 3\log_2 (N^{1/3}) = \log_2 N~\text{qubits},\\
    \texttt{del},\texttt{in\_range} &: 2~\text{qubits},\\
    \text{scratch} &: \text{a small constant number of ancilla}.
\end{align*}
Hence the total ancilla requirement is
\begin{equation}
    a = 2 + \lceil \log_2 D' \rceil + 1 + \log_2 N + 2 + a_{\text{scratch}}.
    \label{eq:ancilla}
\end{equation}

\subsection{Decomposition and Per-Call Gate Counts}

Each call of $U_{G'}$ is composed of:
\begin{itemize}
    \item $n_{\mathrm{small\_adds}}=4$ additions/subtractions acting on $\ell=\log_2(N^{1/3})$-bit registers,
    \item $n_{\mathrm{big\_adds}}=2$ additions/subtractions acting on the full $\log_2 N = 3\ell$-bit registers,
    \item one out-of-range oracle $O_{rg}$,
    \item small overhead from $O_t$, $Z$, and the Hadamard layers,
    \item data loading step
\end{itemize}

\paragraph{Adder gate counts.}
For an $n$-bit addition, the Remaud adder in Remaud `24~\cite{Remaud24} requires exactly
\begin{equation}
    \text{CNOT count} = 14n - 10, \ \
    \text{T-gate count} = 10n - 3.
    \label{eq:remaud_adder}
\end{equation}
Using these formulas, the total adder costs for one block-encoding call are
\begin{align*}
    \text{CNOT:} \quad & 
        n_{\mathrm{small\_adds}}(14\ell - 10) +
        n_{\mathrm{big\_adds}}(14\cdot3\ell - 10) \\&= 140\ell - 60, \\
    \text{T:} \quad & 
        n_{\mathrm{small\_adds}}(10\ell - 3) +
        n_{\mathrm{big\_adds}}(10\cdot3\ell - 3) \\&= 100\ell - 18.
\end{align*}
\paragraph{$O_{rg}$ cost.}
The out-of-range oracle $O_{rg}$ implements five control conditions, four of which flip both the \texttt{del} and \texttt{in\_range} qubits, and one of which flips only \texttt{del} conditioned on \texttt{in\_range}$=0$ and the value of $d^{val}$. We will use the recent result by Silva and Silva~\cite{Silva25} that shows you can implement a $t$-qubit relative phase toffoli gate using $6t+4$ CNOT gates and zero ancilla. Note that the relative phase does not matter as we are only measuring the del and in-range qubits in the standard basis. Some of these multi-controlled toffolis control on $0$ instead of $1$, so these also require two additional X-gates per $0$-ctrl. With this, the formula for the gate count for each condition of the out-of-range oracle, denoted as $O_{rg}^i$ is:
\begin{equation}
    6(\text{\# of gates controlled on}) + 4 + 2 \cdot (\text{\# of $0$-ctrls})
\end{equation}
Since the first 4 conditions are two toffolis one on the del qubit and one on the in-range qubit, they will incur an multiplicative factor of 2. Finally, we get the gate counts for each condition:
\begin{align*}
    O_{rg}^1 &= 2 \cdot  (6\cdot 3 +4+2\cdot 2) = 52 \\
    O_{rg}^2 \text{ and } O_{rg}^3 &= 2 \cdot (6(1/3 \log N + 2) + 4 + 2(1/3 \log N +1)) \\ &= \frac{16}{3}\log N +36 \\
    O_{rg}^4 &= 2 \cdot (6(1/3 \log N + 2) + 4 + 2(1/3 \log N)) \\&= \frac{16}{3}\log N +32 \\ 
    O_{rg}^5 &\leq (D-D') ( 6 (\lceil \log(D-D') \rceil + 1) + 4 \\ & \hspace{6.5em}+ 2 (\lceil \log(D-D') \rceil + 1)) \\& = (D-D') \cdot (8 \lceil \log(D-D') \rceil + 12)
\end{align*}
where the definitions for $D$ and $D'$ can be found in App.~\ref{sec:BE_GW_flow}. Note that for $O_{rg}^5$, it is an inequality because we have to make an assumption on the number of $0$-ctrls as it is an unknown. Then putting everything together, 
\begin{equation}
    O_{rg} \text{ cost } = \frac{32}{3}\log N + (D-D') \cdot (8 \lceil \log(D-D') \rceil + 12) +  84
\end{equation}
\paragraph{$O_t, Z$, and Hadamard layer cost.} The $O_t$ oracle is one toffoli plus two toffolis with a control on $\ket{0}$ instead of on $\ket{1}$. Using the 16 elementary gate toffoli construction mentioned above, we get
\begin{align}
    \text{$O_t$ cost }= 16 + 2\cdot 18 = 52 \text{ 
elementary gates}.
\end{align}
The $Z$-gate is just a single rotation costing $1$ gate. Then, the 2 Hadamard layers each contain $\log (2D)$ Hadamard gates, giving 
\begin{align}
    \text{Hadamard layer cost } = 2 \log(2D) \text{ 
elementary gates}.
\end{align}
\paragraph{Data loading cost.}
From Appendix A of Sunderhauf et al/~\cite{Snderhauf2024}, we find that the data loading step requires $D$ CNOT gates and $D$ singe-qubit rotation gates. This is because we cleverly manipulated $D$ to always be a power of $2$. So with this,
\begin{align}
    \text{Data loading cost }= 2D \text{ 
elementary gates}.
\end{align}
\paragraph{Final gate count.}
Combining everything from the previous sections we get that 
\begin{equation} \label{eq:elemntry_gate_count_UG}
\begin{aligned}
    g_{U_{G'}} \leq & \frac{272}{3}\log N + 2\log(2D) + 2D \\&+ (D-D') \cdot (8 \lceil \log(D-D') \rceil + 12).
\end{aligned}
\end{equation} 
\subsection{Numerical Comparison with Classical Solvers}
\label{sec:quantum_vs_classical}

To assess the practicality of our QLS algorithm, we compare the explicit gate-count estimates in
Sec.~\ref{sec:explicit_counts} with the classical solver benchmarks of
Greer et al.~\cite{greer2022comparison}, who evaluated linear solvers
for three-dimensional discrete fracture network (DFN) models of
subsurface flow.

\paragraph{Classical solver benchmarks.}  
Greer et al.~\cite{greer2022comparison} compared direct and iterative solvers for three‑dimensional discrete fracture network (DFN) flow problems on networks up to the sizes tested in their paper. For the representative test studied here (three fracture sets, 5,512 fractures; conforming mesh with $N=7{,}758{,}411$ nodes and 15,699,327 triangles; minimum feature size $h=0.025$\,m) they report that sparse Cholesky factorization is the fastest method overall, and that conjugate gradient preconditioned by algebraic multigrid (CG+AMG) is the most competitive iterative fallback. On their single‑workstation experiments (128\,GB RAM, Intel i9 class CPU) wall times for the best methods on this problem scale lie in the envelope $\sim 10^{2}$–$10^{3}$ seconds; Cholesky completed the 7.76M‑node case but its memory use approaches practical machine limits as $N$ and factor fill grow.

\paragraph{Quantum resource estimate.}  
Using Dalzell’s QLS bound~\cite{dalzell2024} applied to the explicit block‑encoding $U_{G'}$ (parameters chosen for concreteness: $D=5{,}512$, $D'=\lceil D/2\rceil=2{,}756$, $\kappa=2^{20}$, $\varepsilon=10^{-3}$) we obtain the following conservative logical resources for solving the same linear system with $N=7{,}758{,}411$: the Dalzell query bound evaluates to

\[
Q \approx 1.31\times 10^{8},
\]

the per‑call gate cost from the block‑encoding formulas is

\[
g_{U_{G'}} \approx 3.09\times 10^{5}\ \text{elementary logical gates},
\]

(a dominant contribution from the $(D-D')$ oracle term), a conservative segmented state‑prep model gives

\[
g_{U_b}\approx 1.10\times 10^{4},
\]

and, assuming worst‑case controlled‑call promotion (factor 16) for all queries, the total logical gate budget is

\[
G_{\mathrm{total}}\;\lesssim\;16\,Q\,g_{U_{G'}}+32\,Q\,g_{U_b}\;\approx\;6.9\times 10^{14}.
\]

The register/ancilla accounting yields $n_{\mathrm{logical}}\approx 70$–85 logical qubits (we conservatively quote $\sim80$).

Converting the total logical gate count into a fault-tolerant wall-time estimate requires modeling execution in terms of repeated quantum error-correction (QEC) cycles. Elementary physical gate times on state-of-the-art superconducting hardware are on the order of $t_{\mathrm{phys}}\sim 20$ ns. For example, in Arute et al. \cite{Arute2019}, they achieve a maximum gate speed of $25$ ns. However, in a surface-code architecture, wall-clock time is not set directly by physical gate duration. Instead, execution is synchronized to repeated QEC cycles that interleave entangling gates, stabilizer measurement, resonator reset, readout, and classical processing. Recent experimental demonstrations report QEC cycle times of approximately $t_{\mathrm{cycle}}\approx 1.1\,\mu\mathrm{s}$, together with an average real-time decoder latency of approximately $63\,\mu\mathrm{s}$ \cite{googleQEC2025}. 

Importantly, the decoder latency does not multiply the total number of QEC cycles. Using Pauli-frame tracking, Pauli correction operations are recorded in classical memory rather than physically applied to qubits, allowing syndrome extraction and decoding to proceed in parallel with subsequent QEC cycles \cite{Riesebos_Pauli_Frames}. In this execution model, the decoder latency contributes only an additive overhead rather than a per-cycle multiplicative cost. Consequently, the dominant contribution to wall-clock runtime is the QEC cycle time itself, not the classical decoding latency.

Logical gates must span multiple QEC cycles to suppress logical error rates below threshold. Denoting by $c_{\mathrm{gate}}$ the number of QEC cycles required per logical operation (which depends on code distance and gate type), the effective wall-time cost per logical gate is 
\[
t_{\mathrm{logical}} \approx c_{\mathrm{gate}}\, t_{\mathrm{cycle}}.
\]
Conservatively taking $c_{\mathrm{gate}}\sim 10$, the effective logical gate duration becomes $11\,\mu\mathrm{s}$, despite underlying physical gates being two orders of magnitude faster. Applying this model to our estimated logical gate count $G_{\mathrm{total}}\approx 6.9\times 10^{14}$, the total runtime is on the order of decades on current fault-tolerant superconducting platforms. The additional $\sim 63\,\mu\mathrm{s}$ decoder latency contributes only a constant overhead and does not change this asymptotic estimate.

Several foreseeable advances could significantly reduce this runtime. Architectural improvements such as faster readout, reduced measurement latency, and improved qubit coherence could reduce QEC cycle times from the current $\sim 1\,\mu\mathrm{s}$ toward the $\sim 100$ ns regime. Improved scheduling and parallelism across logical qubits, block-encoding oracles, and state preparation routines could reduce effective circuit depth relative to the total logical gate count. On the algorithmic side, tighter gate analysis and improved QLS constructions with reduced constant factors could decrease $G_{\mathrm{total}}$ by one or more orders of magnitude. Taken together, such developments suggest that cumulative constant-factor reductions of several orders of magnitude in wall time are plausible over longer time horizons. Nevertheless, even under optimistic assumptions, the quantum runtime remains substantially larger than that of classical solvers for the instances considered here. Achieving a practical quantum advantage will therefore require not only hardware and compilation improvements, but also better preconditioning methods to reduce the runtime for large, ill-conditioned linear systems.

\paragraph{Modeling assumptions and limitations.}  
The classical and quantum columns differ in both modeling and resource semantics and these differences shape the comparison. Greer et al.~\cite{greer2022comparison} solved the fracture‑only DFN (flow confined to lower‑dimensional fracture planes), which reduces the degrees of freedom by not considering the underlying rock as being permeable. On the other hand, our quantum model can be adapted to represent more general coupled fracture–matrix operators but at increased oracle complexity (reflected in the $(D-D')$ term). The quantum query complexity scales approximately linearly with $\kappa$ (and logarithmically with $1/\varepsilon$), while the classical direct method (Cholesky) is essentially independent of $\kappa$ for runtime but sensitive to memory and fill‑in. Finally, the quantum column reports logical resources (gates, logical qubits) that must be converted to physical wall time via a fault‑tolerance model; direct wall‑time comparison without that mapping is therefore not meaningful.

\paragraph{Summary.}  
Anchored to the same DFN test ($N=7{,}758{,}411$), Greer et al.\ demonstrate that optimized classical solvers (Cholesky; CG+AMG) solve the problem in minutes on commodity hardware using tens–hundreds of GB of memory. The Dalzell QLS applied to the explicit block‑encoding requires $\mathcal{O}(10^{14}\!-\!10^{15})$ logical gates under conservative assumptions and $\mathcal{O}(10^{1}-10^{2})$ logical qubits. Therefore, the quantum approach offers exponentially compressed memory but incurs a very large logical‑depth cost dominated by $\kappa$ and oracle structure. Practical quantum competitiveness at this problem scale thus requires (i) significant reduction of $\kappa$ (for example by embedding effective preconditioners into the block‑encoding), (ii) oracle compression so $D'\approx D$ (reducing the heavy $(D-D')$ term), and (iii) lower controlled‑call overhead via optimized multi‑control and compilation strategies.

\section{Conclusion}
Finding useful, real-world applications of QLS algorithms remains an outstanding–and daunting–task. In this work, we have demonstrated that many of the critical components required for applying QLS algorithms can be realized efficiently in the case of 3D discretized heterogeneous elliptic partial differential equations, with explicit examples for the case of 3D fracture flow. Preparing the input state $\ket{b}$, which represents a source or boundary condition vector, is often sparse or structured in practice and can therefore be efficiently implemented. The system matrix $G$ is highly structured and often has a nice classical description. Under the assumption that the number of distinct non-zero elements in the matrix is polylogarithmic in $N$, we can block encode $G$ with $O(\text{polylog }N)$ gate complexity. Furthermore, for many applications we are interested in global properties or local observables of our solution state, which can be extracted from $\ket{x}$ with low overhead.
Combining all of these results with recent QLS algorithms~\cite{costa2021optimal, dalzell2024}, solving the system scales as $O(\kappa \log(1/\epsilon)) = O(N^{2/3} \ \text{polylog }N \cdot \log(1/\epsilon))$. This is better than classical runtime of $O(N \log N \cdot \log(1/\epsilon))$, and importantly, it comes with an exponential reduction in memory. This opens the door to gaining a quantum advantage for simulation of three dimensional heterogeneous Poisson PDEs, but much work remains to be done to make this algorithms practically useful.

However, our analysis also uncovers a critical limitation: reducing the effective condition number $\kappa$ remains a major obstacle to achieving more substantial speedups, which may be necessary for this to be useful on future large-scale, fault-tolerant hardware. Despite having a preconditioner that improves the scaling of the condition number, by block encoding the two matrices separately, we can never reduce the asymptotic runtime of QLS algorithms. While we mention several recent works that offer promising ideas for quantum-compatible preconditioning, further analysis is needed to assess their impact in settings like ours. This suggests that reducing the effective condition number is highly restrictive in the quantum context, and must be treated as a significant limitation in any practical QLS application.

Looking ahead, quantum-compatible strategies for mitigating effective condition number scaling are important. Our results highlight how a carefully tailored, interdisciplinary approach bridging quantum algorithms and domain specific structure can result in a quantum advantage, albeit small in this case. Without improved quantum preconditioning methods, the runtime of our algorithm is dominated by the condition number of the system matrix. If the matrix is naturally well conditioned, like in the 3D groundwater flow case, we can achieve a quantum advantage.

\appendix 
\section{Proof $\|G^{-1}\| = O(1)$} \label{sec:Proof of ||G^{-1}||}
To start, since $G$ is positive definite, square, and symmetric, we can say that $\sigma_{min}(G) = \lambda_{min}(G)$ (where $\lambda_{min}(G)$ is the smallest eigenvalue of $G$). With this, the discrete Poincaré inequality in bounded domains with suitable boundary conditions (e.g., Dirichlet) for our equation of interest $Gh = f$ is:
\begin{equation}
    \|h\|_{L^2} \leq C \cdot \|\nabla_d \  h \|_{L^2}.
\end{equation}
Here, $\nabla_d$ is the discrete gradient operator and $C$ is a constant that depends on our chosen boundary conditions and the physical domain we are considering (in this case an $L \times L \times L$ section of material). So, $C$ is a constant that does not grow with the system size $N$. 

We can write the smallest eigenvalue using the Rayleigh quotient as
\begin{equation}
    \lambda_{min}(G) = \min_{u\neq 0} \frac{u^T Gu}{u^Tu}.
\end{equation} 
Now expanding $u^TGu$ gives
\begin{align*}
    u^TGu &= \sum_{a,b=1}^N u_a \cdot G_{ab} \cdot u_b \\
    &= \sum_{a=1}^N u_a^2 \cdot G_{aa} + \sum_{a\neq b} u_a \cdot G_{ab} \cdot u_b\\
    &= \sum_{a=1}^N \sum_{b \in \text{nbhd}(a)} \frac{\K_{ab}}{(\Delta x)^2} \cdot u_a^2 - \sum_{a\neq b} \frac{\K_{ab}}{(\Delta x)^2} \cdot u_a u_b \\
    &= \sum_{\{a,b\} \ \text{neighbors}} \frac{\K_{ab}}{(\Delta x)^2} (u_a^2 -2u_au_b +u_b^2)\\
    &\geq \K_{min} \cdot \sum_{\{a,b\} \ \text{neighbors}} \frac{(u_a - u_b)^2}{(\Delta x)^2} \\
    &= \K_{min} \cdot \| \nabla_d \ u\|_{L^2}^2
\end{align*}
where $\text{nbhd}(a) = \{b \in \{1,\dots,N\} \setminus \{a\} : G_{ab} \neq 0\}$ and $\sum_{\{a,b\} \ \text{neighbors}}$ is a sum is over unordered pairs. So by now applying the discrete Poincaré inequality we get that
\begin{align*}
    \lambda_{min}(G) &= \min_{u\neq 0} \frac{u^T Gu}{u^Tu} \\
    &\geq \frac{\K_{min} \cdot \| \nabla_d \ u\|_{L^2}^2}{u^Tu} \\
    &\geq \frac{\K_{min}}{C^2} \cdot \frac{\|u\|_{L^2}^2}{u^Tu} \\
    &= \frac{\K_{min}}{C^2}
\end{align*}
Now putting everything together, 
$$ \|G^{-1}\| = \frac{1}{\lambda_{min}(G)} \leq \frac{C^2}{\K_{min}} = O(1) $$
since both $C$ and $\K_{min}$ are constants.
\section{Bound on $D$ for groundwater flow}\label{sec:D_bound_gw_flow}
\begin{figure}[H]
    \centering
    \includegraphics[width=\columnwidth]{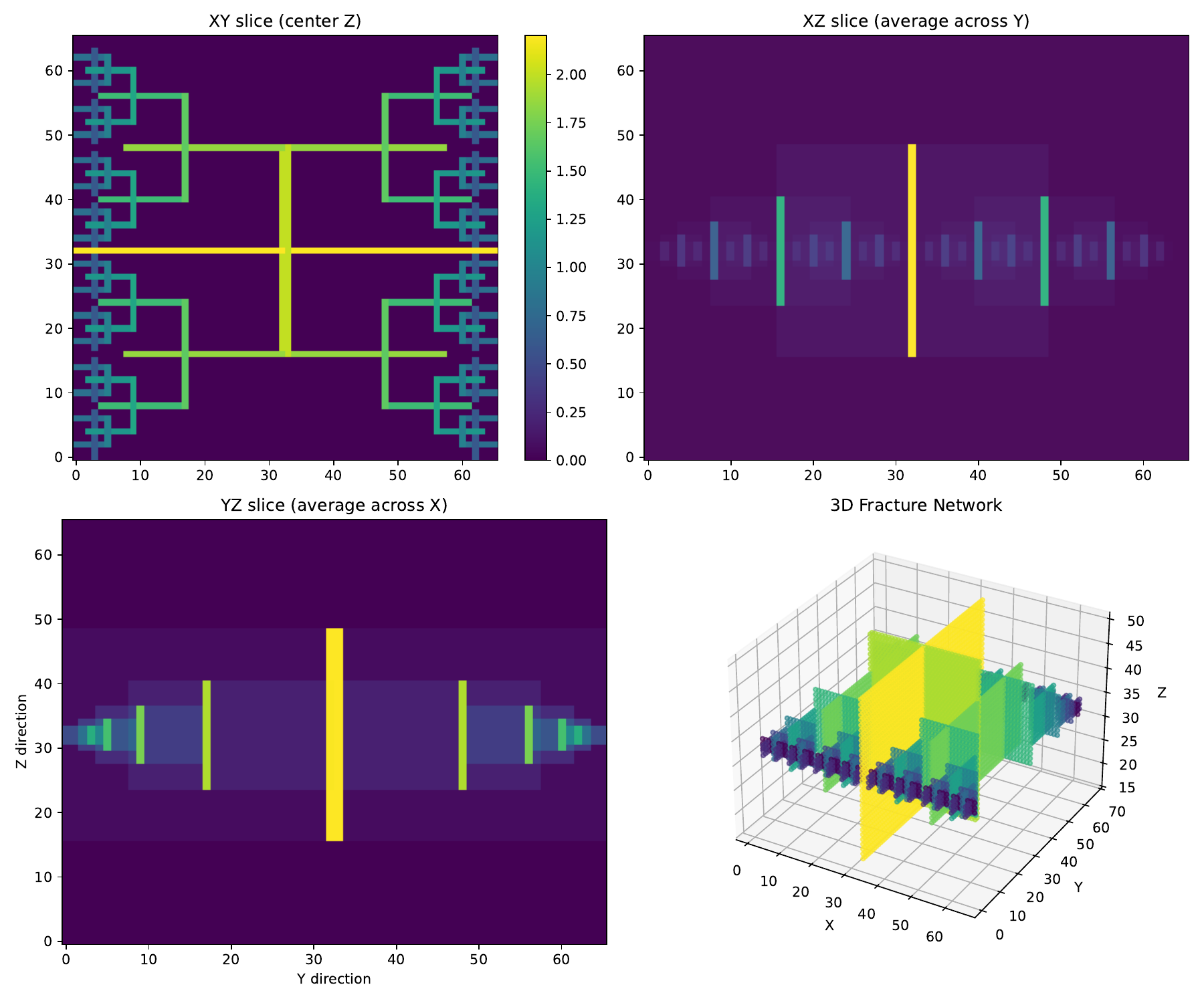}
    \caption{This is an extension of the 2D pitchfork fracture pattern introduced in~\cite{Golden_2022} to 3D. The 2D pitchfork fracture is started in opposite directions of the XY-plane, where each fracture scale is assigned a depth based on its length (the smaller the fracture scale, the smaller the depth). The different colors represent different cell permeability values, and in this particular example, we have $5$ fracture scales.}
	\label{fig:3D_pitchfork}
\end{figure}
In practice, we are discretizing all the fracture scales, which can provide an advantage over classical approaches which truncate the fractures scales and only model large fractures.

\begin{figure}[H]
    \centering
    \includegraphics[width=0.155\textwidth]{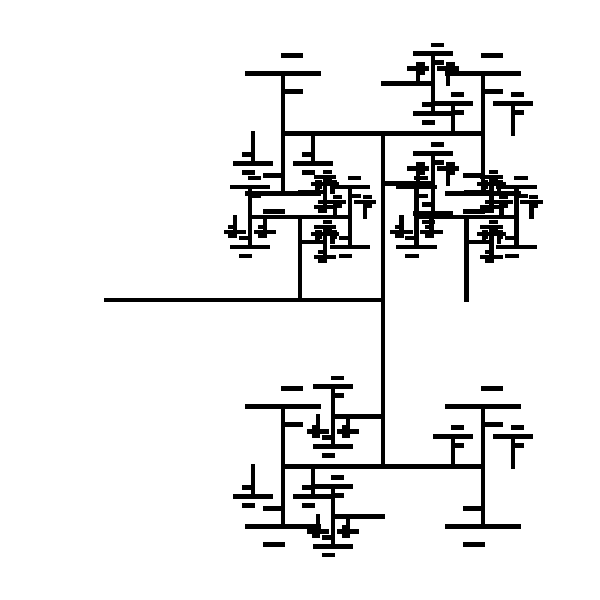}
    \includegraphics[width=0.155\textwidth]{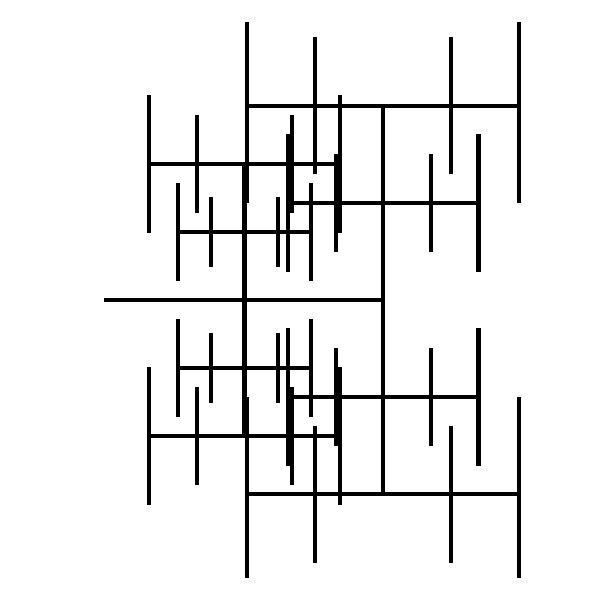}
    \includegraphics[width=0.155\textwidth]{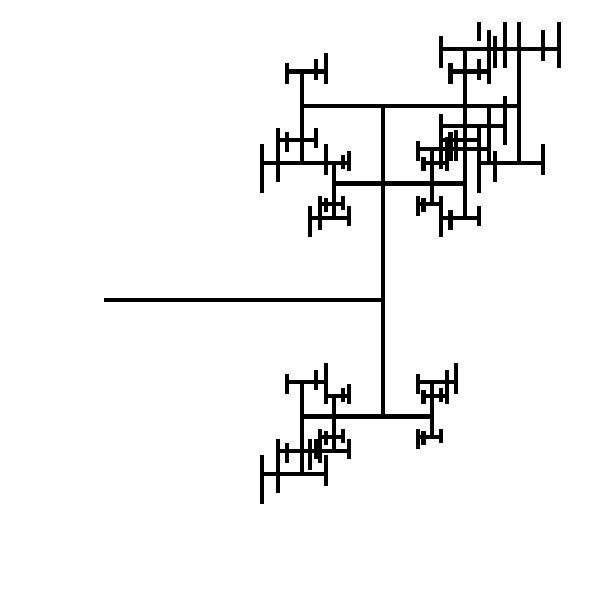}
    \caption{Two dimensional cross sections of different fractal models of fine-grained fracture structure with simple arithmetic formulas. These are examples of fracture structures that could lead to efficient block-encodings for the groundwater flow matrix $G'$.}
	\label{fig:fractals}
\end{figure}

\begin{figure}[H]
    \centering
    \includegraphics[width=\columnwidth]{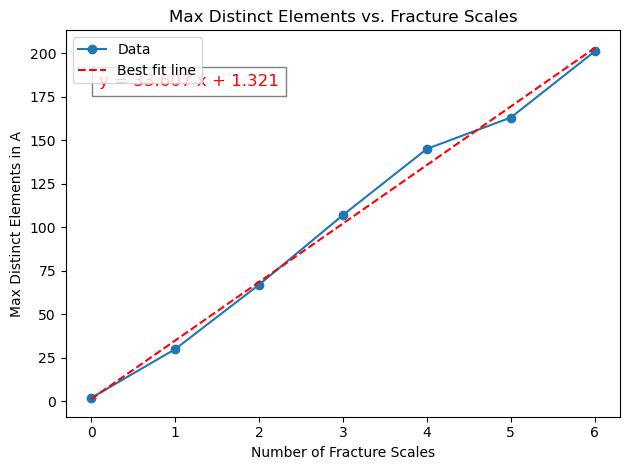}
    \caption{For the 3D pitchfork pattern shown in Fig.~\ref{fig:3D_pitchfork}, we observe that the number of distinct elements in the groundwater flow matrix $G'$ grows roughly linearly with the number of fracture scales $F$. This suggests that the theoretical bound of $O(F^{12})$ from eq.~\ref{eq:distinct_elements_vs_F} may be loose for some practical examples of 3D heterogeneous Poisson problems.}
    \label{fig:linear_fractal_scaling}
\end{figure}

In most classical simulations of groundwater flow, the locations of very large fractures are known \emph{a priori}, and smaller fractures are placed in the grid by sampling from a distribution chosen to match observational data.
Instead we propose using fractal patterns for adding small fractures, for examples see Fig.~\ref{fig:fractals}.
Such fractal patterns, which are based on simple arithmetic formulas and thus can be efficiently encoded in a quantum circuit, could be fine-tuned to match the characteristics and existing data for a given region. In some example fractal patterns following simple arithmetic formulas, we also get significantly better scaling than the theoretical upper bound of $O(F^{12})$. For the 3D pitchfork pattern in Fig \ref{fig:linear_fractal_scaling} we appear to get a linear relationship between the fracture scale and number of distinct elements, which is a significant speed up for practical uses of this algorithm. 
Geologic fracture networks are often modeled as fractals via an effective permeability \cite{davy2006flow,jafari2012estimation}, which loses critical connectivity information that could put the fracture network above or below a percolation threshold.
Our approach overcomes this limitation and represents all the fracture scales explicitly.

So, as long as the fracture pattern has an arithmetic description, and the number of distinct fracture lengths $F$, i.e. distinct values for $\K_{i,j,k}$, scales logarithmically with the number of cells $N$, then the block encoding circuit we showed in App. \ref{sec:full_BE_circuit} gives a block encoding $U_{G'}$ with scaling $O(\log N + D \log D) = O(\text{polylog }N)$.  
\section{Block Encoding $\Delta^{-1}$} \label{sec:BE_inv_laplacian}
The key to constructing an efficient block encoding for $\Delta^{-1}$ is the fast inversion technique introduced by Tong et al.~\cite{tong2021fast}.
Specifically, they give an algorithm for calculating $A^{-1}$ for normal matrices $A$.
This algorithm requires an efficient oracle $O_D$ for a diagonal matrix $D$ and a unitary $V$, where $A= V D V^\dagger$.
Given these then one can calculate $D^{-1}$ and thus $A^{-1}$ using $O_D$, $O_D^\dagger$, $V$, and $V^\dagger$ each exactly once.
In the case of the discrete Laplacian, the necessary oracles for implementing the fast inverse are relatively simple and efficient.
For a 2D discrete Laplacian on a $n \times n$ grid, the eigenvalues are given by
\begin{equation} \label{eq:Laplacian_eigenvalues}
	\lambda_{k_x, k_y} = -\frac{1}{2\sin^2(\frac{\pi}{2(n+1)})} \left( \sin^2\left( \frac{k_x \pi}{2(n+1)} \right) + \sin^2\left( \frac{k_y \pi}{2(n+1)} \right) \right),
\end{equation}
where $k_x, k_y$ are the wave numbers associated with the Fourier modes and the eigenvalues have been rescaled so that $\|\Delta^{-1}\| =1$. 
These eigenvalues can be computed efficiently which means that the cost of implementing the oracle $O_D$ is $O(\log N)$.
In addition, we need access to the unitary matrices $V$ and $V^\dagger$, which in this case correspond to the quantum Fourier transform (QFT) and its inverse.
Since QFT can also be implemented with a gate complexity of $O(\log N)$, the block encoding for the inverse Laplacian $U_{\Delta^{-1}}$ also has total cost of $O(\log N)$.
\section{Scaling in \texorpdfstring{$\epsilon$}{epsilon}} \label{sec:scaling_in_eps}
In order to properly determine the scaling of our QLS algorithm, we must also determine how the necessary accuracy $\epsilon$ scales with $N$. 
Rescaling $G$ and possibly applying preconditioning will not change that the final state prepared by the QLS algorithm:
\begin{equation}
	\ket{x} = \ket{G^{-1}b} = G^{-1}b/\|G^{-1}b\|.
\end{equation}
Ideally, we would choose $\epsilon$ such that it does not change the asymptotic runtime of our algorithm. For example, if we pick $\epsilon = \text{poly} \left(\frac{1}{N} \right)$, then $\log (1/\epsilon) = \text{polylog }N$, giving us a runtime of $O(\kappa \cdot \log (1/\epsilon)) =$ $O(N^{2/3} \text{ polylog }N)$, which aligns with the desired scaling.

\begin{figure}[H]
    \centering
    \includegraphics[width=\columnwidth]{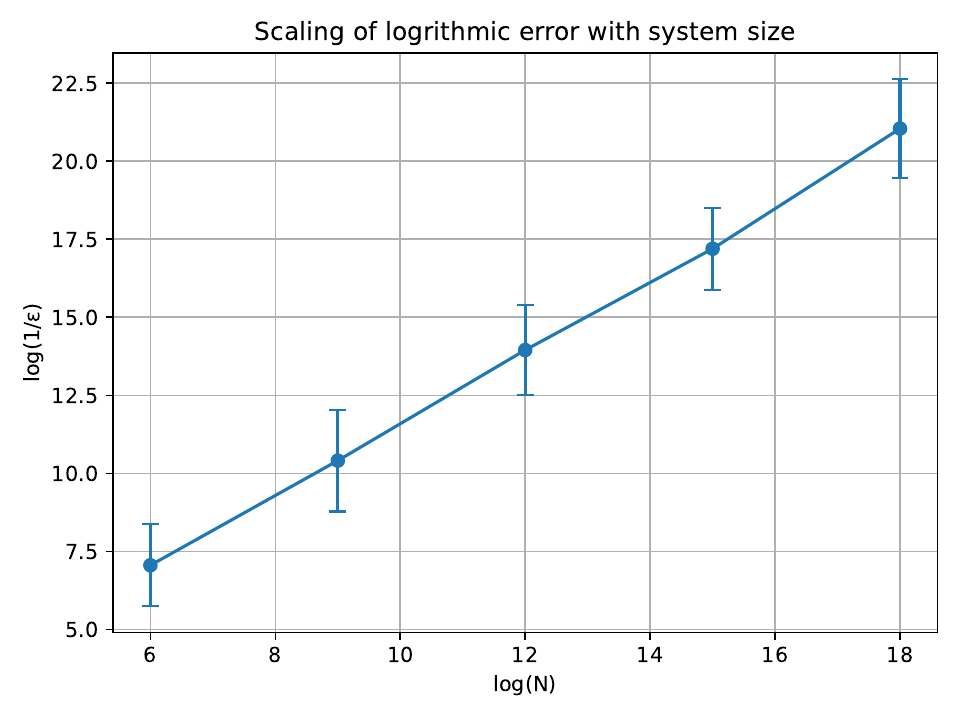}
    \caption{Scaling of the logarithmic error $\log(1/\epsilon)$ with system size $N$. $\epsilon$ is defined as the smallest (non-zero) magnitude of an element in $G^{-1}b/\|G^{-1}b\|$. $G$ is modeled by the modified pitchfork fracture system (seen in Fig. \ref{fig:linear_fractal_scaling}) and $\ket{b}$ represents 20 randomly placed injection/extraction sites for. The error bars represent one standard deviation of results from 100 different choices of $\ket{b}$.}
    \label{fig:scaling_x}
\end{figure}

A natural choice of $\epsilon$, however, is to pick it such that it is on the order of the element of $\ket{x}$ with the smallest non-zero value. This would look like: 
\begin{equation} \label{eq:epsilon_scaling}
	\epsilon = O\left(\frac{\min^{>0}_i\left|(G^{-1} b')_i\right|}{\left\| G^{-1} b' \right\|}\right).
\end{equation}
If $\epsilon$ were larger than this, we lose confidence in the accuracy of the smallest components of the solution, where even the sign could be wrong for a larger epsilon.
This quantity in eq. \ref{eq:epsilon_scaling} is highly problem dependent and depends heavily on the exact form of the decomposition of $b$ in the eigenbasis of $G$. It can, however, be estimated through numerical studies of the system we care about. Numerical studies for groundwater flow, using the 3D pitchfork fractal fracture network shown in Fig.~\ref{fig:3D_pitchfork}, indicate favorable scaling of the required accuracy parameter $\varepsilon$. Using the definition of $\varepsilon$ seen in eq. 27, we observed that $\log(1/\varepsilon)$ scales as $O(\text{polylog}\,N)$ in these simulations. Full results and parameter details are provided in Fig.~\ref{fig:scaling_x}.

\section{Block Encoding $G'$}\label{sec:BE_GW_flow}
\begin{figure}[H]
    \centering
    \includegraphics[width=\columnwidth]{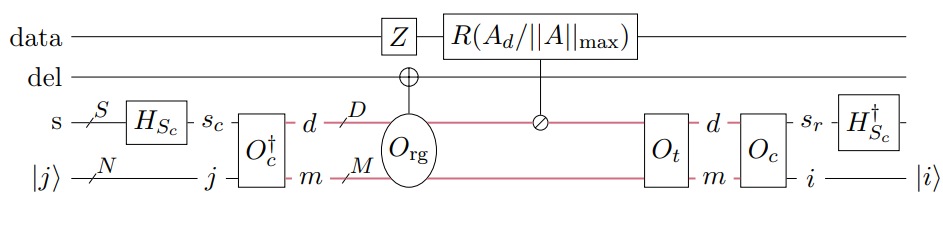}
    \caption{Block encoding circuit for symmetric matrices as given in equation (36) of \cite{Snderhauf2024}. This circuit serves as the template for our block encoding of the discretized 3D heterogeneous Poisson matrix $G'$.}
	\label{fig:sym_BE}
\end{figure}
First, it will be useful to restate some important definitions that will be used frequently throughout this section. \begin{itemize}
    \item $D$ is the number of distinct, non-zero values in $G'$, and $d\in \{0, \ldots, D-1\}$ is a label indicating a specific value 
    \item $M$ is the maximum multiplicity of any non-zero element in $G'$, and $m\in \{0, \ldots M-1\}$ is a label that distinguishes different elements with the same value.
    \item $S$ is the maximum number of nonzero elements per row or column (in the three dimensional groundwater flow case, both the column sparsity and row sparsity are $7$, making $S=7$), and $s_c\in \{0, \ldots, S\}$ indicates a specific non-zero element in a column.  
    \item $i,j \in \{0, \ldots, N\}$ are row and column labels respectively.
\end{itemize}
Now, in addition to $O_c$ and $O_t$, described in section \ref{sec:encode-G}, we also need to define an out of bounds oracle $O_{rg}$ that does the following:
$$ O_{rg} \ket{d}\ket{m}\ket{0} = \begin{cases}
    \ket{d}\ket{m}\ket{0} &\text{if $G'_{i(d,m), j(d,m)} = G'_d$} \\
    \ket{d}\ket{m}\ket{1} &\text{if $G'_{i(d,m), j(d,m)} = 0$}
\end{cases} $$ 
Here, $i(d,m)$ and $j(d,m)$ denote the row and column indices determined by $(d,m)$. Following the symmetric block encoding circuit from \cite{Snderhauf2024}, we additionally require $MD = NS$ to ensure that the block encoding we construct is Hermitian. This can be achieved by padding both $M$ and $S$.

Before we start, we define distinct elements in a way that simplifies the block encoding. There are four distinct sections of $G'$ as given in eq. \ref{eq:G-def}: Section $00$: $a=b$, Section $01$: $b = a \pm 1$, and Section $10$: $b = a \pm N^{1/3}$, and Section $11$: $b= a \pm N^{2/3}$. With this, we label every item as $d = d^{ind} || d^{val}$ where $d^{ind}\in\{0,1\}^2$ is an indicator of which section $d$ is a part of and $d^{val}$ is the label indicating which distinct value $d$ is. From this, we get that if two elements in $G'$, $d_1$ and $d_2$, have the same value but are in different sections, then they will be treated as distinct values under this new labeling.

So, if we let $D^{init}$ be the actual number of distinct elements in $G'$ and  $D'$ be the number of distinct elements after relabeling every $d$ as $d^{ind} || d^{val}$, then we get that $4 \leq D' \leq 4D^{init}$.

With this new labeling scheme, the maximum multiplicity is at most $2(N-1)$ which we will pad to $M= 2N$. Along with that, the number of distinct elements must be a power of $2$ (since we need to be able to represent each distinct element as a binary string), so we will finally let $D = 2^{\lceil\log_2(D')\rceil}$. This gives us $2D \geq 8 >$ max sparsity, which is $7$ in this case. So by padding $S = 2D$, we get $MD = NS = 2ND$ as needed. These extra padded qubits will not create a problem as the out-of-range oracle will handle any cases where our current state does not correspond to a valid non-zero entry in $G'$.  

Now, we are ready to construct $O_c, O_t, $ and $ O_{rg}$. These constructions closely resemble the $2$-dimensional Laplacian example from \cite{Snderhauf2024}, with a couple of changes to deal with the increased number of data elements and slightly different structure. This structure change is due to the fact that $G'$ is closely related to the $3D$ and not $2D$ Laplacian. Once we have created explicit circuits for these oracles, by following Fig. \ref{fig:sym_BE} we will get a block encoding for $G'$ as seen in Fig. \ref{fig:GW_flow}.
\subsection{Transposition Oracle $O_t$} The transposition oracle takes a value $d$ and a multiplicity label $m$ and outputs $(d,m')$ where $m'$ is the element related to $m$ by transposition. Note that there always exists an element $m'$ as our matrix $G'$ is symmetric. \\
Following from \cite{Snderhauf2024} we denote the first bit of $m$ as $m^{hi}$ and the last $\log_2N$ bits as $m^{lo}$. We then let $m^{hi}=0$ denote we are in the lower left triangular matrix (including the diagonal), and let $m^{hi} = 1$ denote we are in the upper triangular matrix. In the lower triangular matrix we choose $m^{lo}$ to be the row index, and in the upper
triangular matrix we choose $m^{lo}$ to be the column index. With this, the transposition oracle will do the following: \begin{align*}
    O_t (d, m^{hi}, m^{lo}) = \begin{cases}
        (d, m^{hi}, m^{lo}) &\text{if $d^{ind} = 00$} \\
        (d, 1-m^{hi}, m^{lo}) &\text{if $d^{ind} \in \{ 01,10, 11\}$}
    \end{cases}
\end{align*}
which as a quantum circuit is
\begin{center}
    \begin{tikzpicture}
\node[scale=0.7] {
    \begin{quantikz}
  \lstick[2]{$d^{ind}$} & & \gate[5]{O_t} & \\
  &  & &\\
  \lstick{$d^{val}$} &  \qwbundle{} &&  \\
  \lstick{$m^{hi}$} & & & \\
  \lstick{$m^{lo}$} & \qwbundle{N} & &
\end{quantikz} = 
    \begin{quantikz}
  \lstick[2]{$d^{ind}$} & & \octrl{1} & \ctrl{1} & \ctrl{1} & \\
  &  & \ctrl{2} & \octrl{2} & \ctrl{2} &\\
  \lstick{$d^{val}$} &  \qwbundle{} && & & \\
  \lstick{$m^{hi}$} & & \targ{} & \targ{}& \targ{} & \\
  \lstick{$m^{lo}$} & \qwbundle{N} & &&&
\end{quantikz}
};
\end{tikzpicture}
\end{center}
Note that the ``/'' symbol on the quantum wires represents that it contains more than one qubit. If there is a number $K$ attached to the ``/'' then that wire contains $\log_2K$ qubits; however, since in the case of $d^{val}$ the size of the channel varies based on the fracture network, we attach no number to it.
\subsection{Column Oracle $O_c$}
The column oracle takes a value $d$ and a number $m$ and outputs the column $j$ and sparsity index $s_c$ corresponding to the $m^{th}$ occurrence of the value $d$.
When we apply the column oracle, we will update the $m^{lo}$ register to be the column $j$ and then concatenate the remaining outputs to make the sparsity index $s_c$. For simplicity, we just describe $O_c: (d,m) \rightarrow j$ below.
\begin{align*}
    O_c (d, m^{hi}, m^{lo}) = \begin{cases}
        m^{lo} &\text{if $d^{ind} = 00$} \\
        m^{lo} &\text{if $m^{hi} = 1$} \\
        m^{lo}-1 &\text{if $m^{hi}=0$ and $d^{ind} = 01$} \\
        m^{lo} - N^{1/3} &\text{if $m^{hi}=0$ and $d^{ind} = 10$} \\
        m^{lo} - N^{2/3} &\text{if $m^{hi}=0$ and $d^{ind} = 11$} 
    \end{cases}
\end{align*}
As a quantum circuit this is:
\begin{center}
    \begin{tikzpicture}
\node[scale=0.7] {
    \begin{quantikz}
  \lstick[2]{$d^{ind}$} &  & \gate[5]{O_c} & \\
  &  & &\\
  \lstick{$d^{val}$} &  \qwbundle{} &&  \\
  \lstick{$m^{hi}$} & & & \\
  \lstick{$m^{lo}$} & \qwbundle{N} & &
\end{quantikz} = 
    \begin{quantikz}
  \lstick[2]{$d^{ind}$} &  & \octrl{1} & \ctrl{1} & \ctrl{1} & \rstick{$s^{hi}$} \\
  &  & \ctrl{2} & \octrl{2} & \ctrl{2} & \rstick{$s^{mid}$}\\
  \lstick{$d^{val}$} &  \qwbundle{} && & &\rstick{$s^{else}$} \\
  \lstick{$m^{hi}$} & & \octrl{1} & \octrl{1}& \octrl{1} & \rstick{$s^{lo}$} \\
  \lstick{$m^{lo}$} & \qwbundle{N} & \gate{-1} & \gate{- N^{1/3}} & \gate{-N^{2/3}} & \rstick{$j$} 
\end{quantikz}
};
\end{tikzpicture}
\end{center}
Here the subtraction gates indicate that we add that amount to the integer representation of $m^{lo}$. So, if we apply the $-1$ gate then $m^{lo} \rightarrow m^{lo} - 1 \pmod{N}$ and similarly for the other two. \\

As mentioned above, and seen in the circuit diagram, the $m^{lo}$ register is updated to represent the column $j$, and the sparsity index is a concatenation of the remaining outputs. In particular, we relabel the outputs of the column oracle in the following way: $s_c = d^{ind} || d^{val} || m^{hi} =s^{hi} || s^{mid} || s^{else} || s^{lo}$. Since $G'$ has at most $7$ elements in each column, we need at most $3$ qubits to represent the sparsity index; however, $s_c$ may contain more than $3$ qubits due to the fact that it was padded. So, we have $s^{hi} || s^{mid} || s^{lo}$ is the part of $s_c$ that actually tells us what the sparsity index is, while $s^{else}$ is simply a byproduct of padding $S$ that has no relevance when deciphering the sparsity index.
\subsection{Out-of-Range Oracle $O_{rg}$}
Since $D$, $M$ and $S$ were padded to ensure $DM = NS$, there are assignments to $(d,m)$ that do not correspond to actual, non-zero values in $G'$ (we refer to these assignments as out-of-range entries and refer to valid assignments to $(d,m)$ as in-range entries). So, out-of-range oracle takes as input $\ket{d}\ket{m}\ket{0}$ and outputs $\ket{d}\ket{m}\ket{1}$ if $(d,m)$ is an out-of-range entry and $\ket{d}\ket{m}\ket{0}$ if $(d,m)$ is an in-range entry. The $0/1$ indicator qubit appended to the end of $\ket{d}\ket{m}$ we label as ``del''. With this, the out-of-range indices $(d,m)= (d^{ind}, d^{val}, m^{hi}, m^{lo})$ are the following:
\begin{enumerate}
    \item $d^{ind} = 00$ and $m^{hi} = 1$
    \item $d^{ind} = 01 $ and $ m^{lo} = 0 \pmod{N^{1/3}}$
    \item $d^{ind}=10 $ and $\left \lfloor \frac{m^{lo} \pmod{N^{2/3}}}{N^{1/3}} \right \rfloor = 0$
    \item $d^{ind} = 11$ and $m^{lo} < N^{2/3}$
    \item $d^{val} > D' $ and $(d^{ind}, m^{hi}, m^{lo})$ does not satisfy any of the above conditions
\end{enumerate}
The constraint on condition five that ``$(d^{ind}, m^{hi}, m^{lo})$ does not satisfy any of the above conditions,'' is added so all five conditions are dependent on each other. We need to ensure this so there is not a case where we accidentally flip the delete qubit twice.

To simplify the implementation of the quantum circuit, we split $m^{lo}$ into three registers $N_a$, $N_b$, and $N_c$, all with $\log_2 N^{1/3}$ qubits. We define these registers such that if $N_a= a$, $N_b=b$ and $N_c = c$ for any $a,b,c \in \{0,1\}^{\log_2 N^{1/3}}$, then we can write $m^{lo} = a+ N^{1/3} b + N^{2/3} c$. With this, the quantum circuit for $O_{rg}$ is the following:
\begin{center}
    \begin{tikzpicture}
\node[scale=0.7] {
    \begin{quantikz}
    \lstick{del} & & \targ{} & \\
  \lstick[2]{$d^{ind}$} &  & \gate[7]{O_{rg}} \wire[u][1]{q}
& \\
  &  & &\\
  \lstick{$d^{val}$} &  \qwbundle{} &&  \\
  \lstick{$m^{hi}$} & & & \\
  \lstick[3]{$m^{lo}$} & \qwbundle{N_c} & &\\
  & \qwbundle{N_b} && \\
  & \qwbundle{N_a} &&
\end{quantikz} = 
    \begin{quantikz}
    \lstick{del} & \targ{} & \targ{} & \targ{} & \targ{} & \targ{} & \\
    \lstick{in range} & \targ{} & \targ{} & \targ{} & \targ{} & \octrl{-1} & \\
  \lstick[2]{$d^{ind}$} & \octrl{-2} & \octrl{-2} & \ctrl{-2} & \ctrl{-2} & & \\
  & \octrl{-1} & \ctrl{-1} & \octrl{-1} & \ctrl{-1} && \\
  \lstick{$d^{val}$} & & & & & \dprime \wire[u][3]{q} & \\
  \lstick{$m^{hi}$} & \ctrl{-2} &&&&& \\
  \lstick[3]{$m^{lo}$} & \qwbundle{N_c}&&&\octrl{-3}&& \\
  & \qwbundle{N_b} && \octrl{-4} &&& \\
  & \qwbundle{N_a} & \octrl{-5} &&&&
\end{quantikz}
};
\end{tikzpicture}
\end{center}
Here, the circle containing $>D'$ (which we will refer to as control-$D'$) is a sequence of control gates that indicates whether $d^{val}> D'$.
Also, notice that we added another qubit labeled ``in range'' that indicates if $(d^{ind}, m^{hi}, m^{lo})$ is a valid in-range entry or not. We then control on this register not being flipped to satisfy the second constraint on condition five.

The construction for control-$D'$ is dependent on the structure of the fracture network (e.g., the number of fracture scales), and thus cannot be explicitly described here; however, the gate complexity of control-$D'$ will be the number of out-of-range values, which is $D- D' = 2^{\lceil \log_2(D') \rceil} - D' = O(D')$. Each gate will also only need to control on $O(\log_2 D')$ qubits as we only need to consider operations on the last $\lceil\log_2(D- D')\rceil$ bits of the input. So, the total gate complexity of the out-of-range oracle is $O(D' \log_2D')$.

\subsection{Full Block Encoding Circuit} \label{sec:full_BE_circuit}

We now have all of the tools we need to block encode $G'$. By using the symmetric block encoding circuit in \cite{Snderhauf2024} (seen in Fig. \ref{fig:sym_BE}) and the three oracles defined above, the full quantum circuit for fracture flow matrix is seen in Fig. \ref{fig:GW_flow}.

Note that the notation of a slash in the control on $R(G'_{d})$ indicates that the gate controls on several values. In particular, it applies a rotation according to what the d-register of the circuit is. Additionally, the template in Fig. \ref{fig:sym_BE} requires the rotation gate to be of the form $R(A_d / \|A\|_{max})$ (where $\|\cdot \|_{max}$ is the maximum absolute value of any element in $A$). The actual requirement is that for all elements $A_{ij}$ in $A$, they must satisfy $A_{ij}\in [-1,1]$. We already satisfy this constraint by our definition of $G'$. Using eq. \ref{eq:bound_on_G'_norm}, we have that for all elements $G_{ij}'$ in $G'$, $|G_{ij}'| \leq \| G' \| \leq 1$. So, it suffices to just rotate about $G'_d$. Finally, the gate complexity is dominated by the additions and out-of-range oracle. The additions require $O(\log_2N + \log_2 N^{1/3})$ gates and the out-of-range oracle requires $O(D' \log_2D')$ gates, giving a total gate complexity of $O(\log_2 N + D' \log_2D')= O(\log_2 N + D^{init} \log_2D^{init})$.

All constructions above assume that we can efficiently identify the value of any matrix element. That is, given a position label $(i,j)$ or an index pair $(d,m)$, we must be able to determine the corresponding entry $G'_{ij}$ and associate it with the correct value label $d \in \{0, \ldots, D-1\}$. This requires an efficient mapping from matrix coordinates (or multiplicity index) to value class, and we assume that this mapping is computable in polylogarithmic time with respect to $N$ and $D$. In practice, this can be implemented either by precomputing a lookup table for all nonzero entries or by using a deterministic rule based on the underlying physical model. For example, the fractal fracture idea mentioned in App. \ref{sec:D_bound_gw_flow} gives an efficient way to do this for groundwater flow.

\onecolumngrid
\begin{figure}[H]
\centering
\begin{tikzpicture}
\node[scale=0.7] {
    \begin{quantikz}
    \lstick{data} &&&&&&&& \gate{Z} &&& \gate{R(G'_{d}) } &&&&&&&& \\
    \lstick{del} & &&&&& \targ{} & \targ{} & \targ{} & \targ{} & \targ{} && &&&&&&& \\
    \lstick{in range} &  &&&&& \targ{} & \targ{} & \targ{} & \targ{} & \octrl{-1} &&&&&&&&&\\
  \lstick{$s^{hi}$} & \gate[4]{H^{\bigoplus \log_2 (2D)}} & \ctrl{1} & \ctrl{1} & \octrl{1} & \midstick[2]{$d^{ind}$} &  \octrl{-2} & \octrl{-2} & \ctrl{-2} & \ctrl{-2} & & \dctrlslash{} & \octrl{1} & \ctrl{1} & \ctrl{1} & \octrl{1} & \ctrl{1} & \ctrl{1} & \gate[4]{H^{\bigoplus \log_2 (2D)}}&\\
  \lstick{$s^{mid}$} & & \ctrl{2} & \octrl{2} & \ctrl{2} && \octrl{-1} & \ctrl{-1} & \octrl{-1} & \ctrl{-1} && \dctrlslash{} \wire[u][1]{q} & \ctrl{2}& \octrl{2} & \ctrl{2} & \ctrl{2} &\octrl{2} & \ctrl{2}&&\\
  \lstick{$s^{else}$} & \qwbundle{} & &&& \midstick{$d^{val}$} & & & & & \dprime \wire[u][3]{q} & \dctrlslash{} \wire[u][5]{q}&&&&&&&&\\
  \lstick{$s^{lo}$} & & \octrl{2} & \octrl{3} & \octrl{1} & \midstick{$m^{hi}$} & \ctrl{-2} &&&&&& \targ{} & \targ{}& \targ{} & \octrl{1} & \octrl{3} & \octrl{2} &&\\
  \lstick[3]{$j$} & \qwbundle{N_c} &&& \gate[3]{+1}& \midstick[3]{$m^{lo}$}&&  && \octrl{-3} &&&&&&\gate[3]{-1} &&&&\rstick[3]{i} \\
  & \qwbundle{N_b} & \gate{+1} &&  & & && \octrl{-4} &&&&&&& & & \gate{-1} && \\
  & \qwbundle{N_a} && \gate{+1} && & & \octrl{-5} &&&&&&&&&\gate{-1}&&&
\end{quantikz}
};
\end{tikzpicture}
\parbox{\textwidth}{%
\caption{Full block encoding for the 3D heterogeneous Poisson equation, following the technique of block encoding symmetric matrices seen in Fig. \ref{fig:sym_BE}.}
\label{fig:GW_flow} }
\end{figure}
\twocolumngrid

\begin{acknowledgements}
 AP and DO gratefully acknowledge support from the Department of Energy, Office of Science, Office of Basic Energy Sciences, Geoscience Research program under Award Number LANLECA1. DO and JG acknowledge support from Los Alamos National Laboratory's Laboratory Directed Research and Development program through project 20220077ER. This work was supported by the U.S. Department of Energy through the Los Alamos National Laboratory, LA-UR-22-24431. Los Alamos National Laboratory is operated by Triad National Security, LLC, for the National Nuclear Security Administration of U.S. Department of Energy (Contract No. 89233218CNA000001).
\end{acknowledgements}

\section*{Data Availability Statement}
The code and fracture network data that support the findings of this study are openly available in GitHub at Ref.~\cite{githubcode}.

\bibliographystyle{apsrev4-2}
\bibliography{references}

\begin{thebibliography}{25}%
\makeatletter
\providecommand \@ifxundefined [1]{%
 \@ifx{#1\undefined}
}%
\providecommand \@ifnum [1]{%
 \ifnum #1\expandafter \@firstoftwo
 \else \expandafter \@secondoftwo
 \fi
}%
\providecommand \@ifx [1]{%
 \ifx #1\expandafter \@firstoftwo
 \else \expandafter \@secondoftwo
 \fi
}%
\providecommand \natexlab [1]{#1}%
\providecommand \enquote  [1]{``#1''}%
\providecommand \bibnamefont  [1]{#1}%
\providecommand \bibfnamefont [1]{#1}%
\providecommand \citenamefont [1]{#1}%
\providecommand \href@noop [0]{\@secondoftwo}%
\providecommand \href [0]{\begingroup \@sanitize@url \@href}%
\providecommand \@href[1]{\@@startlink{#1}\@@href}%
\providecommand \@@href[1]{\endgroup#1\@@endlink}%
\providecommand \@sanitize@url [0]{\catcode `\\12\catcode `\$12\catcode `\&12\catcode `\#12\catcode `\^12\catcode `\_12\catcode `\%12\relax}%
\providecommand \@@startlink[1]{}%
\providecommand \@@endlink[0]{}%
\providecommand \url  [0]{\begingroup\@sanitize@url \@url }%
\providecommand \@url [1]{\endgroup\@href {#1}{\urlprefix }}%
\providecommand \urlprefix  [0]{URL }%
\providecommand \Eprint [0]{\href }%
\providecommand \doibase [0]{https://doi.org/}%
\providecommand \selectlanguage [0]{\@gobble}%
\providecommand \bibinfo  [0]{\@secondoftwo}%
\providecommand \bibfield  [0]{\@secondoftwo}%
\providecommand \translation [1]{[#1]}%
\providecommand \BibitemOpen [0]{}%
\providecommand \bibitemStop [0]{}%
\providecommand \bibitemNoStop [0]{.\EOS\space}%
\providecommand \EOS [0]{\spacefactor3000\relax}%
\providecommand \BibitemShut  [1]{\csname bibitem#1\endcsname}%
\let\auto@bib@innerbib\@empty
\bibitem [{\citenamefont {Costa}\ \emph {et~al.}(2022)\citenamefont {Costa}, \citenamefont {An}, \citenamefont {Sanders}, \citenamefont {Su}, \citenamefont {Babbush},\ and\ \citenamefont {Berry}}]{costa2021optimal}%
  \BibitemOpen
  \bibfield  {author} {\bibinfo {author} {\bibfnamefont {P.~C.}\ \bibnamefont {Costa}}, \bibinfo {author} {\bibfnamefont {D.}~\bibnamefont {An}}, \bibinfo {author} {\bibfnamefont {Y.~R.}\ \bibnamefont {Sanders}}, \bibinfo {author} {\bibfnamefont {Y.}~\bibnamefont {Su}}, \bibinfo {author} {\bibfnamefont {R.}~\bibnamefont {Babbush}},\ and\ \bibinfo {author} {\bibfnamefont {D.~W.}\ \bibnamefont {Berry}},\ }\href {https://doi.org/10.1103/PRXQuantum.3.040303} {\bibfield  {journal} {\bibinfo  {journal} {PRX Quantum}\ }\textbf {\bibinfo {volume} {3}},\ \bibinfo {pages} {040303} (\bibinfo {year} {2022})}\BibitemShut {NoStop}%
\bibitem [{\citenamefont {Dalzell}(2024)}]{dalzell2024}%
  \BibitemOpen
  \bibfield  {author} {\bibinfo {author} {\bibfnamefont {A.}~\bibnamefont {Dalzell}},\ }\href {https://doi.org/10.48550/arXiv.2406.12086} {\bibfield  {journal} {\bibinfo  {journal} {arXiv preprint arXiv:2406.12086}\ } (\bibinfo {year} {2024})}\BibitemShut {NoStop}%
\bibitem [{\citenamefont {Tong}\ \emph {et~al.}(2021)\citenamefont {Tong}, \citenamefont {An}, \citenamefont {Wiebe},\ and\ \citenamefont {Lin}}]{tong2021fast}%
  \BibitemOpen
  \bibfield  {author} {\bibinfo {author} {\bibfnamefont {Y.}~\bibnamefont {Tong}}, \bibinfo {author} {\bibfnamefont {D.}~\bibnamefont {An}}, \bibinfo {author} {\bibfnamefont {N.}~\bibnamefont {Wiebe}},\ and\ \bibinfo {author} {\bibfnamefont {L.}~\bibnamefont {Lin}},\ }\href {https://doi.org/10.1103/PhysRevA.104.032422} {\bibfield  {journal} {\bibinfo  {journal} {Phys. Rev. A}\ }\textbf {\bibinfo {volume} {104}},\ \bibinfo {pages} {032422} (\bibinfo {year} {2021})}\BibitemShut {NoStop}%
\bibitem [{\citenamefont {Costa}\ \emph {et~al.}(2025)\citenamefont {Costa}, \citenamefont {An}, \citenamefont {Babbush},\ and\ \citenamefont {Berry}}]{Costa2025discreteadiabatic}%
  \BibitemOpen
  \bibfield  {author} {\bibinfo {author} {\bibfnamefont {P.~C.}\ \bibnamefont {Costa}}, \bibinfo {author} {\bibfnamefont {D.}~\bibnamefont {An}}, \bibinfo {author} {\bibfnamefont {R.}~\bibnamefont {Babbush}},\ and\ \bibinfo {author} {\bibfnamefont {D.}~\bibnamefont {Berry}},\ }\href {https://doi.org/10.22331/q-2025-10-20-1887} {\bibfield  {journal} {\bibinfo  {journal} {{Quantum}}\ }\textbf {\bibinfo {volume} {9}},\ \bibinfo {pages} {1887} (\bibinfo {year} {2025})}\BibitemShut {NoStop}%
\bibitem [{\citenamefont {Low}\ and\ \citenamefont {Su}(2024)}]{lowsu24qls}%
  \BibitemOpen
  \bibfield  {author} {\bibinfo {author} {\bibfnamefont {G.~H.}\ \bibnamefont {Low}}\ and\ \bibinfo {author} {\bibfnamefont {Y.}~\bibnamefont {Su}},\ }\href {https://arxiv.org/abs/2410.18178v1} {\bibfield  {journal} {\bibinfo  {journal} {arXiv preprint arXiv:2410.18178}\ } (\bibinfo {year} {2024})}\BibitemShut {NoStop}%
\bibitem [{\citenamefont {Deiml}\ and\ \citenamefont {Peterseim}(2024)}]{deiml2024fem}%
  \BibitemOpen
  \bibfield  {author} {\bibinfo {author} {\bibfnamefont {M.}~\bibnamefont {Deiml}}\ and\ \bibinfo {author} {\bibfnamefont {D.}~\bibnamefont {Peterseim}},\ }\href {https://arxiv.org/abs/2403.19512} {\bibfield  {journal} {\bibinfo  {journal} {arXiv preprint arXiv:2403.19512}\ } (\bibinfo {year} {2024})}\BibitemShut {NoStop}%
\bibitem [{\citenamefont {Orsucci}\ and\ \citenamefont {Dunjko}(2021)}]{orsucci2021qls}%
  \BibitemOpen
  \bibfield  {author} {\bibinfo {author} {\bibfnamefont {D.}~\bibnamefont {Orsucci}}\ and\ \bibinfo {author} {\bibfnamefont {V.}~\bibnamefont {Dunjko}},\ }\href {https://arxiv.org/abs/2101.11868} {\bibfield  {journal} {\bibinfo  {journal} {arXiv preprint arXiv:2101.11868}\ } (\bibinfo {year} {2021})}\BibitemShut {NoStop}%
\bibitem [{\citenamefont {Lapworth}\ and\ \citenamefont {S\"underhauf}(2025)}]{LapworthSunderhauf2025}%
  \BibitemOpen
  \bibfield  {author} {\bibinfo {author} {\bibfnamefont {L.}~\bibnamefont {Lapworth}}\ and\ \bibinfo {author} {\bibfnamefont {C.}~\bibnamefont {S\"underhauf}},\ }\href@noop {} {\bibfield  {journal} {\bibinfo  {journal} {arXiv preprint arXiv:2502.20908}\ } (\bibinfo {year} {2025})}\BibitemShut {NoStop}%
\bibitem [{\citenamefont {Strang}(2007)}]{strang2007computational}%
  \BibitemOpen
  \bibfield  {author} {\bibinfo {author} {\bibfnamefont {G.}~\bibnamefont {Strang}},\ }\href@noop {} {\emph {\bibinfo {title} {Computational science and engineering}}},\ \bibinfo {number} {Sirsi i9780961408817}\ (\bibinfo {year} {2007})\BibitemShut {NoStop}%
\bibitem [{\citenamefont {Hyman}\ \emph {et~al.}(2015)\citenamefont {Hyman}, \citenamefont {Karra}, \citenamefont {Makedonska}, \citenamefont {Gable}, \citenamefont {Painter},\ and\ \citenamefont {Viswanathan}}]{hyman2015dfnWorks}%
  \BibitemOpen
  \bibfield  {author} {\bibinfo {author} {\bibfnamefont {J.~D.}\ \bibnamefont {Hyman}}, \bibinfo {author} {\bibfnamefont {S.}~\bibnamefont {Karra}}, \bibinfo {author} {\bibfnamefont {N.}~\bibnamefont {Makedonska}}, \bibinfo {author} {\bibfnamefont {C.~W.}\ \bibnamefont {Gable}}, \bibinfo {author} {\bibfnamefont {S.~L.}\ \bibnamefont {Painter}},\ and\ \bibinfo {author} {\bibfnamefont {H.~S.}\ \bibnamefont {Viswanathan}},\ }\href {https://doi.org/https://doi.org/10.1016/j.cageo.2015.08.001} {\bibfield  {journal} {\bibinfo  {journal} {Computers \& Geosciences}\ }\textbf {\bibinfo {volume} {84}},\ \bibinfo {pages} {10} (\bibinfo {year} {2015})}\BibitemShut {NoStop}%
\bibitem [{\citenamefont {Golden}\ \emph {et~al.}(2022)\citenamefont {Golden}, \citenamefont {O’Malley},\ and\ \citenamefont {Viswanathan}}]{Golden_2022}%
  \BibitemOpen
  \bibfield  {author} {\bibinfo {author} {\bibfnamefont {J.}~\bibnamefont {Golden}}, \bibinfo {author} {\bibfnamefont {D.}~\bibnamefont {O’Malley}},\ and\ \bibinfo {author} {\bibfnamefont {H.}~\bibnamefont {Viswanathan}},\ }\bibfield  {journal} {\bibinfo  {journal} {Scientific Reports}\ }\textbf {\bibinfo {volume} {12}},\ \href {https://doi.org/10.1038/s41598-022-25727-9} {10.1038/s41598-022-25727-9} (\bibinfo {year} {2022})\BibitemShut {NoStop}%
\bibitem [{\citenamefont {Harrow}\ \emph {et~al.}(2009)\citenamefont {Harrow}, \citenamefont {Hassidim},\ and\ \citenamefont {Lloyd}}]{harrow2009quantum}%
  \BibitemOpen
  \bibfield  {author} {\bibinfo {author} {\bibfnamefont {A.~W.}\ \bibnamefont {Harrow}}, \bibinfo {author} {\bibfnamefont {A.}~\bibnamefont {Hassidim}},\ and\ \bibinfo {author} {\bibfnamefont {S.}~\bibnamefont {Lloyd}},\ }\href {https://doi.org/10.1103/PhysRevLett.103.150502} {\bibfield  {journal} {\bibinfo  {journal} {Physical review letters}\ }\textbf {\bibinfo {volume} {103}},\ \bibinfo {pages} {150502} (\bibinfo {year} {2009})}\BibitemShut {NoStop}%
\bibitem [{\citenamefont {Henderson}\ \emph {et~al.}(2024)\citenamefont {Henderson}, \citenamefont {Kath}, \citenamefont {Golden}, \citenamefont {Percus},\ and\ \citenamefont {O'Malley}}]{henderson2023addressingquantumsfineprint}%
  \BibitemOpen
  \bibfield  {author} {\bibinfo {author} {\bibfnamefont {J.~M.}\ \bibnamefont {Henderson}}, \bibinfo {author} {\bibfnamefont {J.}~\bibnamefont {Kath}}, \bibinfo {author} {\bibfnamefont {J.~K.}\ \bibnamefont {Golden}}, \bibinfo {author} {\bibfnamefont {A.~G.}\ \bibnamefont {Percus}},\ and\ \bibinfo {author} {\bibfnamefont {D.}~\bibnamefont {O'Malley}},\ }\href {https://doi.org/10.1038/s41598-024-52759-0} {\bibfield  {journal} {\bibinfo  {journal} {Scientific Reports}\ }\textbf {\bibinfo {volume} {14}},\ \bibinfo {pages} {3592} (\bibinfo {year} {2024})}\BibitemShut {NoStop}%
\bibitem [{\citenamefont {Sünderhauf}\ \emph {et~al.}(2024)\citenamefont {Sünderhauf}, \citenamefont {Campbell},\ and\ \citenamefont {Camps}}]{Snderhauf2024}%
  \BibitemOpen
  \bibfield  {author} {\bibinfo {author} {\bibfnamefont {C.}~\bibnamefont {Sünderhauf}}, \bibinfo {author} {\bibfnamefont {E.}~\bibnamefont {Campbell}},\ and\ \bibinfo {author} {\bibfnamefont {J.}~\bibnamefont {Camps}},\ }\href {https://doi.org/10.22331/q-2024-01-11-1226} {\bibfield  {journal} {\bibinfo  {journal} {Quantum}\ }\textbf {\bibinfo {volume} {8}},\ \bibinfo {pages} {1226} (\bibinfo {year} {2024})}\BibitemShut {NoStop}%
\bibitem [{\citenamefont {Hyman}\ \emph {et~al.}(2016)\citenamefont {Hyman}, \citenamefont {Aldrich}, \citenamefont {Viswanathan}, \citenamefont {Makedonska},\ and\ \citenamefont {Karra}}]{hyman2016fracture}%
  \BibitemOpen
  \bibfield  {author} {\bibinfo {author} {\bibfnamefont {J.}~\bibnamefont {Hyman}}, \bibinfo {author} {\bibfnamefont {G.}~\bibnamefont {Aldrich}}, \bibinfo {author} {\bibfnamefont {H.}~\bibnamefont {Viswanathan}}, \bibinfo {author} {\bibfnamefont {N.}~\bibnamefont {Makedonska}},\ and\ \bibinfo {author} {\bibfnamefont {S.}~\bibnamefont {Karra}},\ }\href@noop {} {\bibfield  {journal} {\bibinfo  {journal} {Water Resources Research}\ }\textbf {\bibinfo {volume} {52}},\ \bibinfo {pages} {6472} (\bibinfo {year} {2016})}\BibitemShut {NoStop}%
\bibitem [{\citenamefont {Lubinski}\ \emph {et~al.}(2023)\citenamefont {Lubinski}, \citenamefont {Johri}, \citenamefont {Varosy}, \citenamefont {Coleman}, \citenamefont {Zhao}, \citenamefont {Necaise}, \citenamefont {Baldwin}, \citenamefont {Mayer},\ and\ \citenamefont {Proctor}}]{Lubinski23}%
  \BibitemOpen
  \bibfield  {author} {\bibinfo {author} {\bibfnamefont {T.}~\bibnamefont {Lubinski}}, \bibinfo {author} {\bibfnamefont {S.}~\bibnamefont {Johri}}, \bibinfo {author} {\bibfnamefont {P.}~\bibnamefont {Varosy}}, \bibinfo {author} {\bibfnamefont {J.}~\bibnamefont {Coleman}}, \bibinfo {author} {\bibfnamefont {L.}~\bibnamefont {Zhao}}, \bibinfo {author} {\bibfnamefont {J.}~\bibnamefont {Necaise}}, \bibinfo {author} {\bibfnamefont {C.}~\bibnamefont {Baldwin}}, \bibinfo {author} {\bibfnamefont {K.}~\bibnamefont {Mayer}},\ and\ \bibinfo {author} {\bibfnamefont {T.}~\bibnamefont {Proctor}},\ }\href {https://doi.org/10.1109/TQE.2023.3253761} {\bibfield  {journal} {\bibinfo  {journal} {IEEE Transactions on Quantum Engineering}\ }\textbf {\bibinfo {volume} {PP}},\ \bibinfo {pages} {1} (\bibinfo {year} {2023})}\BibitemShut {NoStop}%
\bibitem [{\citenamefont {Remaud}(2024)}]{Remaud24}%
  \BibitemOpen
  \bibfield  {author} {\bibinfo {author} {\bibfnamefont {M.}~\bibnamefont {Remaud}},\ }in\ \href {https://doi.org/10.1145/3665870.3665875} {\emph {\bibinfo {booktitle} {Proceedings of Recent Advances in Quantum Computing and Technology}}},\ \bibinfo {series and number} {ReAQCT '24}\ (\bibinfo  {publisher} {Association for Computing Machinery},\ \bibinfo {address} {New York, NY, USA},\ \bibinfo {year} {2024})\ p.\ \bibinfo {pages} {56–61}\BibitemShut {NoStop}%
\bibitem [{\citenamefont {Silva}\ and\ \citenamefont {da~Silva}(2025)}]{Silva25}%
  \BibitemOpen
  \bibfield  {author} {\bibinfo {author} {\bibfnamefont {J.~D.~S.}\ \bibnamefont {Silva}}\ and\ \bibinfo {author} {\bibfnamefont {A.~J.}\ \bibnamefont {da~Silva}},\ }\href@noop {} {\  (\bibinfo {year} {2025})},\ \Eprint {https://arxiv.org/abs/2507.00400} {arXiv:2507.00400 [quant-ph]} \BibitemShut {NoStop}%
\bibitem [{\citenamefont {Greer}\ \emph {et~al.}(2022)\citenamefont {Greer}, \citenamefont {Hyman},\ and\ \citenamefont {O'Malley}}]{greer2022comparison}%
  \BibitemOpen
  \bibfield  {author} {\bibinfo {author} {\bibfnamefont {S.}~\bibnamefont {Greer}}, \bibinfo {author} {\bibfnamefont {J.}~\bibnamefont {Hyman}},\ and\ \bibinfo {author} {\bibfnamefont {D.}~\bibnamefont {O'Malley}},\ }\bibfield  {journal} {\bibinfo  {journal} {Water Resources Research}\ }\href {https://doi.org/10.1029/2021WR031188} {10.1029/2021WR031188} (\bibinfo {year} {2022})\BibitemShut {NoStop}%
\bibitem [{\citenamefont {Arute}\ \emph {et~al.}(2019)\citenamefont {Arute}, \citenamefont {Arya}, \citenamefont {Babbush} \emph {et~al.}}]{Arute2019}%
  \BibitemOpen
  \bibfield  {author} {\bibinfo {author} {\bibfnamefont {F.}~\bibnamefont {Arute}}, \bibinfo {author} {\bibfnamefont {K.}~\bibnamefont {Arya}}, \bibinfo {author} {\bibfnamefont {R.}~\bibnamefont {Babbush}}, \emph {et~al.},\ }\href {https://doi.org/10.1038/s41586-019-1666-5} {\bibfield  {journal} {\bibinfo  {journal} {Nature}\ }\textbf {\bibinfo {volume} {574}},\ \bibinfo {pages} {505} (\bibinfo {year} {2019})}\BibitemShut {NoStop}%
\bibitem [{\citenamefont {{Google Quantum AI}}\ and\ \citenamefont {Collaborators}(2025)}]{googleQEC2025}%
  \BibitemOpen
  \bibfield  {author} {\bibinfo {author} {\bibnamefont {{Google Quantum AI}}}\ and\ \bibinfo {author} {\bibnamefont {Collaborators}},\ }\href {https://doi.org/10.1038/s41586-024-08449-y} {\bibfield  {journal} {\bibinfo  {journal} {Nature}\ }\textbf {\bibinfo {volume} {638}},\ \bibinfo {pages} {920} (\bibinfo {year} {2025})}\BibitemShut {NoStop}%
\bibitem [{\citenamefont {Riesebos}\ \emph {et~al.}(2017)\citenamefont {Riesebos}, \citenamefont {Fu}, \citenamefont {Varsamopoulos}, \citenamefont {Almudever},\ and\ \citenamefont {Bertels}}]{Riesebos_Pauli_Frames}%
  \BibitemOpen
  \bibfield  {author} {\bibinfo {author} {\bibfnamefont {L.}~\bibnamefont {Riesebos}}, \bibinfo {author} {\bibfnamefont {X.}~\bibnamefont {Fu}}, \bibinfo {author} {\bibfnamefont {S.}~\bibnamefont {Varsamopoulos}}, \bibinfo {author} {\bibfnamefont {C.~G.}\ \bibnamefont {Almudever}},\ and\ \bibinfo {author} {\bibfnamefont {K.}~\bibnamefont {Bertels}},\ }in\ \href {https://doi.org/10.1145/3061639.3062300} {\emph {\bibinfo {booktitle} {Proceedings of the 54th Annual Design Automation Conference 2017}}},\ \bibinfo {series and number} {DAC '17}\ (\bibinfo  {publisher} {Association for Computing Machinery},\ \bibinfo {address} {New York, NY, USA},\ \bibinfo {year} {2017})\BibitemShut {NoStop}%
\bibitem [{\citenamefont {Davy}\ \emph {et~al.}(2006)\citenamefont {Davy}, \citenamefont {Bour}, \citenamefont {De~Dreuzy},\ and\ \citenamefont {Darcel}}]{davy2006flow}%
  \BibitemOpen
  \bibfield  {author} {\bibinfo {author} {\bibfnamefont {P.}~\bibnamefont {Davy}}, \bibinfo {author} {\bibfnamefont {O.}~\bibnamefont {Bour}}, \bibinfo {author} {\bibfnamefont {J.-R.}\ \bibnamefont {De~Dreuzy}},\ and\ \bibinfo {author} {\bibfnamefont {C.}~\bibnamefont {Darcel}},\ }\href@noop {} {\bibfield  {journal} {\bibinfo  {journal} {Geological Society, London, Special Publications}\ }\textbf {\bibinfo {volume} {261}},\ \bibinfo {pages} {31} (\bibinfo {year} {2006})}\BibitemShut {NoStop}%
\bibitem [{\citenamefont {Jafari}\ and\ \citenamefont {Babadagli}(2012)}]{jafari2012estimation}%
  \BibitemOpen
  \bibfield  {author} {\bibinfo {author} {\bibfnamefont {A.}~\bibnamefont {Jafari}}\ and\ \bibinfo {author} {\bibfnamefont {T.}~\bibnamefont {Babadagli}},\ }\href@noop {} {\bibfield  {journal} {\bibinfo  {journal} {Journal of Petroleum Science and Engineering}\ }\textbf {\bibinfo {volume} {92}},\ \bibinfo {pages} {110} (\bibinfo {year} {2012})}\BibitemShut {NoStop}%
\bibitem [{\citenamefont {Pechan}(2024)}]{githubcode}%
  \BibitemOpen
  \bibfield  {author} {\bibinfo {author} {\bibfnamefont {A.}~\bibnamefont {Pechan}},\ }\href {https://github.com/Austin-Pechan/Block-encoding-the-3D-heterogeneous-Poisson-equation-with-application-to-fracture-flow} {\bibinfo {title} {Code for: Block encoding the 3d heterogeneous poisson equation with application to fracture flow}},\ \bibinfo {howpublished} {\url{https://github.com/Austin-Pechan/Block-encoding-the-3D-heterogeneous-Poisson-equation-with-application-to-fracture-flow}} (\bibinfo {year} {2024})\BibitemShut {NoStop}%
\end{thebibliography}%
\end{document}